\documentclass[12pt,preprint]{aastex}

\usepackage{afterpage}

\input epsf

\renewcommand{\ion}[2]{#1\,{\sc #2}}
\newcommand{\al} {$\alpha$}
\newcommand{\be} {$\beta$}
\newcommand{\gam}{$\gamma$}
\newcommand{\lam}{$\lambda$}

\newcommand{\ecs} {erg\,cm$^{-2}$\,s$^{-1}$} 
\newcommand{\ecsa} {erg cm$^{-2}$ s$^{-1}$ \AA$^{-1}$} 
\newcommand{\kms} {km\ s$^{-1}$} 
\newcommand{\deltE}{\Delta\kern-1ptE}
\newcommand{\iue}{\emph{IUE}}
\newcommand{\fuse}{\emph{FUSE}}

\begin{document}

\title{FUSE OBSERVATIONS OF THE SYMBIOTIC STAR AG DRACONIS}
 
\author{P.R.\ Young,\altaffilmark{1,2}
A.K.\ Dupree,\altaffilmark{1}
B.R.\ Espey,\altaffilmark{3}
S.J.\ Kenyon,\altaffilmark{1} \and
T.B.\ Ake,\altaffilmark{4,5}
}

\altaffiltext{1}{Smithsonian Astrophysical Observatory, 60
Garden Street, Cambridge, MA 02138}

\altaffiltext{2}{Present address: Rutherford Appleton Laboratory,
  Chilton, Didcot, 
  Oxfordshire, OX11 0QX, U.K.}

\altaffiltext{3}{Trinity College Dublin, Dublin 2, Ireland}

\altaffiltext{4}{Department of Physics and Astronomy, Johns Hopkins
University, 3400 North Charles Street, Baltimore, MD 21218}

\altaffiltext{5}{Computer Science Corporation, Lanham, MD 20706}

\begin{abstract}
Spectra of the bright symbiotic star AG Draconis
(BD+67$^{\circ}$922) in the wavelength range 905--1187\,\AA\ obtained
with the \emph{Far Ultraviolet Spectroscopic Explorer} (\fuse) are
presented. The spectra show a number of narrow, nebular emission lines,
together with a uniform continuum from the hot component of the
system, and numerous interstellar absorption lines.
We infer the existence of \ion{Ne}{viii} in the AG Dra
nebula through the identification of the \ion{Ne}{vii} 973.3\,\AA\
recombination line.
The emission line spectrum is dominated by intense lines of
\ion{O}{vi}, but also shows weaker lines from highly-ionized
ions including \ion{Ne}{v}, \ion{Ne}{vi}, \ion{S}{iv},
\ion{S}{vi}. Members of 
the \ion{He}{ii} Balmer series can be identified up to n=20.
Lines of \ion{Fe}{ii} and \ion{Fe}{iii} fluoresced by \ion{O}{vi}
\lam1032 are identified at wavelengths
1141.172\,\AA\ and 1142.429\,\AA, respectively.
The emission lines are shown to be produced in a plasma with an
electron temperature of 20--30,000~K, photoionized
by the white dwarf. The
\ion{Ne}{vi} \lam997/\lam999 ratio shows that this ion and all others
except perhaps \ion{Ne}{vii} are formed at least 300 white dwarf radii
from the white dwarf.
Revised wavelengths for the \ion{Ne}{v} $2s^22p^2$ $^3P_{0,1}$ --
$2s2p^3$ $^5S_2$ and \ion{Ne}{vi} $2s^22p$ $^2P_J$ -- $2s2p^2$
$^4P_{J^\prime}$ 
transitions are published.

\end{abstract}


\section{Introduction}\label{sect.intro}

Symbiotic stars are binary stars that are identified
spectroscopically as typically having three components: a giant, a hot
star (usually a white dwarf), and a circumstellar nebula, ionized by
the hot component's radiation field.  Accretion of material from the
giant's wind onto the white dwarf can lead to eruptions that may last
several years and raise the system's brightness by several magnitudes.
AG Draconis (BD+67$^{\circ}$922) is one of the brightest symbiotics at
UV and  X-ray wavelengths and undergoes outbursts in 3--6 year long
intervals that repeat every 15 years or so. The  observations
described here took place during quiescence.  The components of AG Dra
are a K0--4 giant and a  hot star, most likely a white dwarf
\citep{miko95}. For convenience we will refer to the hot component as
a white dwarf for the remainder of this article. The temperature of
the white dwarf  is derived through fitting a blackbody curve to the
continuum observed at X-ray and UV wavelengths, although contrasting
results are found: \citet{greiner97} finding a temperature of
170,000~K from ROSAT X-ray spectra, while \citep{miko95} finding
100,000~K from IUE spectra.

The system is a non-eclipsing binary with an orbital period of
$\approx$550~days \citep{mein79, galis99, fekel00}. The radial velocity was
found by \citet{demed99} to be $-147.46\pm 3.54$~\kms\ and they give
the rotational velocity of the giant as $5.9\pm 1.0$~\kms. The
abundances of the giant are well determined \citep{smith96} and
demonstrate that it is metal deficient and possibly a barium star.
The distance to AG Dra is crucial in fixing many of the properties of
the system, yet it is rather uncertain, with the \emph{Hipparcos}
satellite only yielding a lower limit of 1~kpc. \citet{miko95} argue
from a classification of the giant for a distance of 2.5~kpc, which is
the value we adopt in the present work.

AG Dra is a particularly attractive target in the far ultraviolet on
account of the low extinction \citep[E$_{B-V}$=0.05][]{miko95} along
its sightline, 
resulting from the high latitude of the system ($b=41^\circ$). Spectra
from the International Ultraviolet Explorer (\iue) showed strong
emission lines from highly ionized species including \ion{He}{ii} \lam1640, \ion{C}{iv}
\lam\lam1548,1550 and \ion{N}{v} \lam1238,1242 \citep{viotti83}. The
first reported 
observations for wavelengths below 1200\,\AA\ were made by
\citet{schmid99} who identified strong emission lines from
\ion{O}{vi} \lam1032,1038 and \ion{He}{ii} \lam1085.

We report here observations made with the \emph{Far Ultraviolet
Spectroscopic Explorer} (\fuse) in the wavelength region
905--1187\,\AA\ made in 2000 March and 2001 April during a quiescent
period of the system. These spectra are at a higher resolution and
sensitivity than any previous spectra in this range and reveal a
number of new plasma diagnostics available for interpreting the nebular
emission from the system.

\section{The instrument}

\fuse\ is described in detail in \citet{moos00} and \citet{blair01}, and we describe
the main features of the spectrograph here. There are four telescope
channels labeled LiF1, LiF2, SiC1, and SiC2, each containing a mirror, a focal plane assembly and a
grating. The labels refer to the coatings on the mirrors and gratings,
either silicon carbide (SiC) or lithium fluoride (LiF). The LiF coatings are
highly efficient at wavelengths above $\approx 1000$\,\AA, but have
very little efficiency for smaller wavelengths, hence the SiC channels
extend the \fuse\ wavelength range down to 900\,\AA.

There are two detectors, with the LiF1 and SiC1 spectra being imaged
on detector 1, and LiF2 and SiC2 imaged on detector 2. Each detector
consists of two segments, labeled A and B, which are read out
independently by the detector electronics. The LiF and SiC spectra on
each segment are treated individually by the \fuse\ calibration
pipeline, and so one refers to, e.g., the LiF1A spectrum, the SiC2B
spectrum, etc. As there is a physical gap between each detector's two
segments, 
there is a gap of $\approx 10$\,\AA\ between the A and B spectra for
each channel.

There are three apertures on the focal plane assembly that are
available for observing, the largest ($30^{\prime\prime}\times
30^{\prime\prime}$) and smallest ($1.5^{\prime\prime}\times
20^{\prime\prime}$) apertures were used for the AG Dra observations.

Guiding is done via a fine error sensor, which is fixed to the LiF1
channel. The target can move in the remaining channel apertures due to
thermally-induced mirror motions. The motion of the SiC channels is
generally greater than that for LiF2 and occasionally the target can
leave the aperture.

\section{Data reduction}

Version 1.8.7 of the \fuse\ calibration pipeline was run on the raw data
files. In particular, versions 6 and 10 of the flux and calibration
files were used, respectively. For the SiC spectra, only that portion of the
observation for which the target was in the aperture (see
Sect.~\ref{sect.obs}) was run through the pipeline to minimize
background counts in the spectra.

\section{Observations}\label{sect.obs}

The components of AG Dra are not resolved by \fuse\ and so the spectra
presented here show emission from the entire system.
Two observations were performed through programs S312 and
P248, and details are given in Table~\ref{agdra-obs}. The SiC channels
were only aligned for small portions of the total exposure time in
2000~March, leading to effective exposure times of 440~s and 300~s for
SiC1 and SiC2, respectively.
The \ion{O}{vi}
\lam1032 flux from the 2000~March observation was $2.4\times
10^{-10}$~\ecsa\ in the center of the line, exceeding the \fuse\ flux
limit of $1.0\times 
10^{-10}$~\ecsa, and so a special observing
sequence was employed for the second observation to ensure the safety
of the \fuse\ detectors. 
AG Dra
was first observed through the 1.5$^{\prime\prime}$ $\times$
20$^{\prime\prime}$ slit (HIRS) which has a throughput around 50\% of that of
the standard 30$^{\prime\prime}$ $\times$
30$^{\prime\prime}$ slit and so the count rate on the detector at
\lam1032 was reduced. Coupled with the expected fall in the \lam1032
flux due to observing at a different orbital phase of the system
(Sect.~\ref{sect.time-var}), the effective \lam1032 flux then fell below the \fuse\ flux limit.
Alignment of the target in the  aperture is much more
difficult with the HIRS slit and so observation P2480101 consisted of
three individual exposures. The \lam1032 line was used to check target
alignment during the exposures as it is observed in all four of the
\fuse\ telescope channels and has a high count rate.
The LiF1 channel was aligned throughout,
with only a small loss of flux in the third exposure. LiF2 showed a
loss of flux of around 10\% compared to LiF1. A significant number of
counts were only detected from SiC1 in the third exposure, but there
are rapid variations in the count rate during this exposure and so
these data are not useful. SiC2 showed a relatively uniform \lam1032
count rate during exposures 2 and 3, but very few counts in exposure
1.

\begin{deluxetable}{lllllll}
\tablecaption{Observing parameters for AG Dra observation\label{agdra-obs}}
\tablehead{Date        &Phase\tablenotemark{a} &Dataset ID &Exposure 
  &Aperture &Time  &Exposure time (s)}
\startdata
2000 March 16 &0.540 &S31202 &1 &LWRS &15:57 &2388 \\
2001 April 25 &0.277 &P24801 &1 &HIRS &05:28 &901 \\
            &&&2 &HIRS &05:50 &788 \\
            &&&3 &HIRS &06:09 &451 \\
            &&P24802 &1 &LWRS &06:57 &2060 \\
\enddata
\tablenotetext{a}{Using ephemeris of \citet{fekel00}.}
\end{deluxetable}

The loss of flux in the
SiC channels was anticipated and so a second observation was obtained with AG
Dra in the LWRS aperture, but with the LiF focal plane assemblies moved
such that the 
target would not lie in the aperture. The effective areas of the SiC
channels are lower than those of the LiFs, while the degree of
astigmatism is greater and so the count rate per unit area on the
detector is much lower. This results in the count rate in the \lam1032
line lying at a safe level. No loss of flux is evident during the single
SiC-only exposure. A weak signal from the \ion{O}{vi} lines was
detected in the LiF channels due to scattered light from AG Dra.

\begin{figure}
\epsscale{0.9}
\plotone{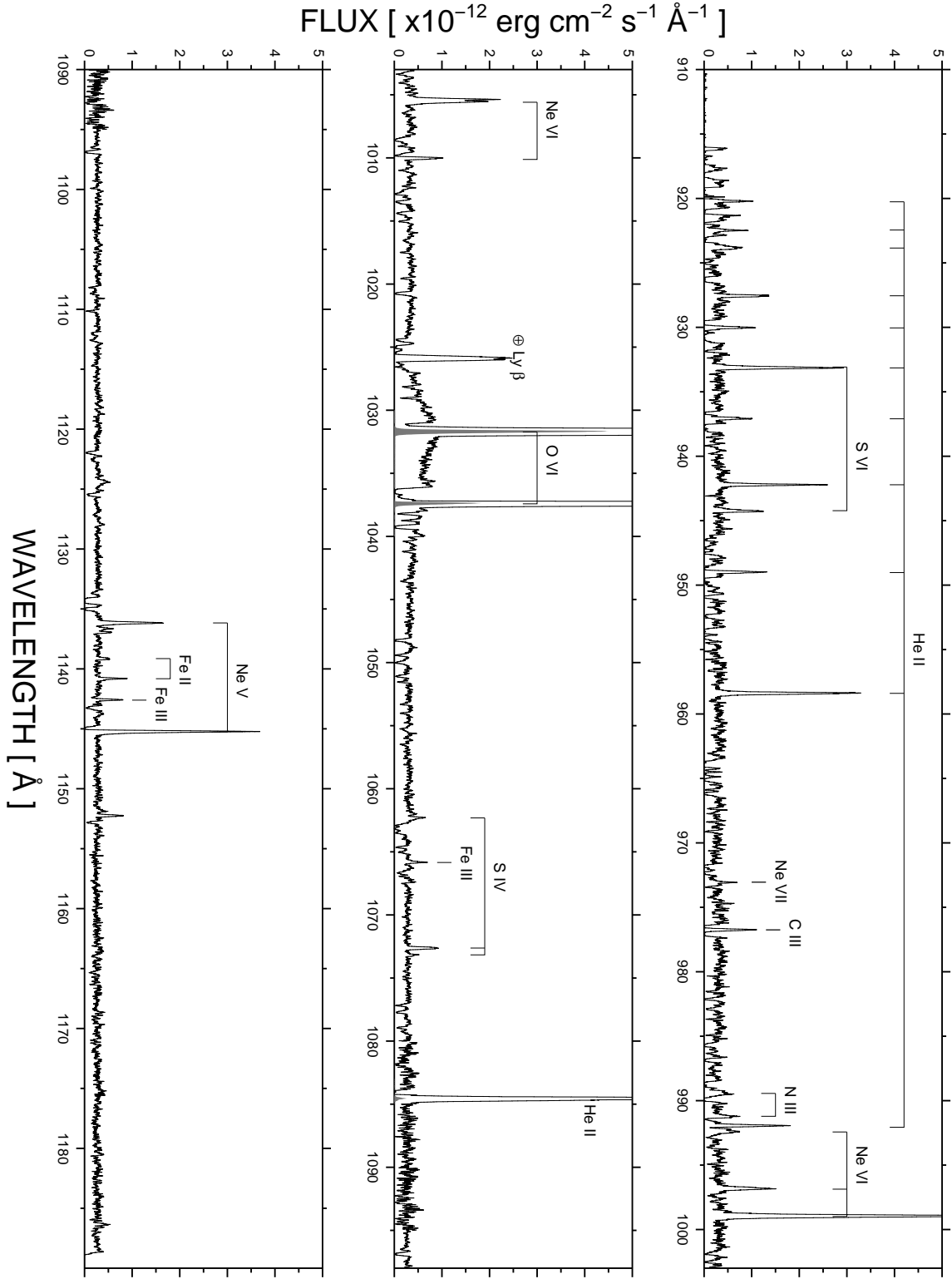}
\caption{\fuse\ spectra of AG Dra from SiC1B (top), LiF1A (middle) and
LiF2B (bottom). Spectra from SiC1A (1080--1090\,\AA) and SiC2B
(1085--1095\,\AA) have been used to complete the wavelength
coverage. The shaded profiles in the middle plot show the \ion{O}{vi}
and \ion{He}{ii} lines reduced by a factor of 50.}
\label{fig.spectra}
\end{figure}

\section{The Spectrum and Line List}

Fig.~\ref{fig.spectra} displays the spectrum of AG Dra over the
complete \fuse\ wavelength range assembled from the LiF spectra of the
2000 March observation and the SiC spectra of the 2001 April
observation. The key features are (i) a uniform continuum from the
white dwarf; (ii) narrow, nebular emission lines from highly-ionized
species; and (iii) interstellar absorption lines.

The present work will largely focus on the emission lines in the
spectrum. We briefly discuss some properties of the interstellar lines
here. Comparisons of the absorption line wavelengths with those of the
emission lines shows that they are redshifted relative to the radial
velocity of the system by $\approx 110$~\kms\ and so are from the
interstellar medium. Absorption by the hydrogen Lyman series,
\ion{O}{i}, \ion{N}{i}, \ion{N}{ii}, \ion{Fe}{ii}, \ion{Si}{ii} and
\ion{Ar}{i} is seen, as well as numerous lines of H$_2$.
Only one component to the absorption lines is seen; in particular no H$_2$
absorption is seen at the radial velocity of the star. In the outer
layers of the star one may expect the AG Dra nebula to cool
sufficiently to form a significant layer of H$_2$. We estimate an
upper limit of $10^{16}$~cm$^{-2}$ to the H$_2$ column density of this layer
from our non-detection.

The interstellar H$_2$ absorption spectrum shows no lines from levels
with $J^\prime\ge 5$, indicating relatively low excitation
temperatures ($<1000$~K). The H$_2$ lines serve as useful fiducial marks for
fixing the wavelength scale of the \fuse\ spectra as many of the lines
are not saturated (unlike the atomic species). We adopt a velocity of
$-23$~\kms\ for the interstellar lines based on \iue\ spectra
\citep{viotti83}.

In the following sections some references are made to the interstellar
H$_2$ lines, and we make use of the shorthand notation for referring
to the H$_2$ transitions introduced by \citet{barnstedt00}. Thus,
e.g.,  L13R3 refers to the $\nu^\prime=13$ to
$\nu^{\prime\prime}=0$, R(3) transition of the H$_2$ Lyman band; while
W0R3 refers to the $\nu^\prime=0$ to
$\nu^{\prime\prime}=0$, R(3) transition of the Werner band.

\subsection{Line list}

The only AG Dra emission lines previously reported in the \fuse\
wavelength band have been the \ion{O}{vi} \lam\lam1032, 1038 lines, and
\ion{He}{ii} \lam1085 \citep{schmid99}. The complete list of lines found
in the \fuse\ spectrum is given in Table~\ref{tbl.lines}.

The measured wavelengths given for each line are those measured
through Gaussian fits to the spectral lines, corrected to be
placed in a reference frame where the interstellar absorption lines
have a velocity of $-23$~\kms\ -- see Sect.~\ref{line-wid}. The
spectrum with the largest effective area at the particular wavelength
considered was used to derive the wavelengths. The LiF measurements
(990--1187\,\AA) 
are all from the 2000~March observation, while the SiC measurements
(905--1095\,\AA) 
are from the 2001~April observation.
Two fluxes are given for each line, corresponding to the two
measurement dates. The first SiC measurements are more uncertain due
to short exposure times.

Line identifications have been based on solar spectra, in particular,
the catalogs of \citet{feldman97} and \citet{curdt01}. There are
several lines in the AG Dra spectrum for which no obvious candidates
from the solar spectra could be found. Some of these lines are
identified here and are marked with a ($*$) -- further details can be
found in Sect.~\ref{sect.fluor}.

\begin{deluxetable}{lllllll}
\tablecaption{AG Dra emission lines in the \fuse\ waveband.\label{tbl.lines}}
\tabletypesize{\scriptsize}
\tablehead{
\lam$_{\rm meas}$$^a$ &\multicolumn{2}{l}{Flux$^b$ ($10^{-14}$\,\ecs)}  
    &\lam$_{\rm rest}$ &Ion &Transition &Wavelength\\
(\AA)   &Mar.~2000 &Apr.~2001&(\AA) &&&Reference \\}
\startdata
920.082  &-- &7&920.561 &\ion{He}{ii} &Balmer (n=20) &1 \\
922.336  &-- &7&922.748 &\ion{He}{ii} &Balmer (n=18) &1\\
923.674  &-- &10 &924.147 &\ion{He}{ii} &Balmer (n=17)&1 \\
927.411 &18 &18 &927.851 &\ion{He}{ii} &Balmer (n=15) &1 \\
929.886 &   &20 &930.342 &\ion{He}{ii} &Balmer (n=14) &1 \\
932.989 &64 &48 &933.378  &\ion{S}{vi}
         &$3s$ $^2S_{1/2}$ -- $3p$ $^2P_{3/2}$ &2 \\
         &   &   &933.449 &\ion{He}{ii} &Balmer (n=13) &1\\
936.929 &--   &11 &937.394 &\ion{He}{ii} &Balmer (n=12) &1 \\
942.068 &35 &33 &942.513 &\ion{He}{ii} &Balmer (n=11) &1 \\
944.100 &19 &15 &944.523 &\ion{S}{vi}
         &$3s$ $^2S_{1/2}$ -- $3p$ $^2P_{3/2}$ &2 \\
948.843 &--   &15 &949.329 &\ion{He}{ii}
         &Balmer (n=10) &1 \\
958.241 &68 &56 &958.698 &\ion{He}{ii} &Balmer (n=9) &1\\
972.917 &-- &4  &973.35  &\ion{Ne}{vii} 
         &$2s~2p$ $^1P_1$ -- $2p^2$ $^1D_2$ &3 \\
976.603 &9 &8 &977.020 &\ion{C}{iii} 
         &$2s^2$ $^1S_0$ -- $2s~2p$ $^1P_1$ &4 \\
989.388  &-- &9&989.787  &\ion{N}{iii}
         &$2s^2~2p$ $^2P_{1/2}$ -- $2s~2p^2$ $^2D_{3/2}$ &2 \\
991.078  &--&7 &991.564  &\ion{N}{iii} 
         &$2s^2~2p$ $^2P_{3/2}$ -- $2s~2p^2$ $^2D_{5/2}$ &2 \\
991.795  &--&24 &992.363 &\ion{He}{ii} &Balmer (n=7) &1 \\
992.298  &18 &14&992.731 &\ion{Ne}{vi} 
         &$2s^2~2p$ $^2P_{1/2}$ -- $2s~2p^2$ $^4P_{3/2}$ &5\\
996.671  &29 &26&997.169 &\ion{Ne}{vi} 
         &$2s^2~2p$ $^2P_{1/2}$ -- $2s~2p^2$ $^4P_{1/2}$ &5\\
998.812 &140 &135 &999.291 &\ion{Ne}{vi} 
         &$2s^2~2p$ $^2P_{3/2}$ -- $2s~2p^2$ $^4P_{5/2}$ &5\\
1005.323$^c$ &60 &52 &1005.789 &\ion{Ne}{vi}
        &$2s^2~2p$ $^2P_{3/2}$ -- $2s~2p^2$ $^4P_{3/2}$ &5 \\
1009.899  &12 &12 &1010.323 &\ion{Ne}{vi}
        &$2s^2~2p$ $^2P_{3/2}$ -- $2s~2p^2$ $^4P_{1/2}$ &5 \\
1031.549  &5290 &3930 &1031.926 &\ion{O}{vi} 
        &$2s$ $^2S_{1/2}$ -- $2p$ $^2P_{3/2}$ &6\\
1037.262  &1750 &1340 &1037.617  &\ion{O}{vi} 
        &$2s$ $^2S_{1/2}$ -- $2p$ $^2P_{3/2}$ &6\\
1062.213 &4 &-- &1062.664 &\ion{S}{iv}
         &$3s^2~3p$ $^2P_{1/2}$ -- $3s~3p^2$ $^2D_{3/2}$ &2 \\
1065.707 &5 &6 &1066.195 &\ion{Fe}{iii} 
         &$3d^6$ $^3G_5$ -- $3d^5~(^4G)4p$ $^3F_4$  ($*$) &7\\
1072.527 &13 &6 &1072.973 &\ion{S}{iv} 
         &$3s^2~3p$ $^2P_{3/2}$ -- $3s~3p^2$ $^2D_{5/2}$ &2\\
1073.068 &3 &5 &1073.518 &\ion{S}{iv} 
         &$3s^2~3p$ $^2P_{3/2}$ -- $3s~3p^2$ $^2D_{3/2}$ &2\\
1084.004 &240 &229 &1084.940 &\ion{He}{ii} &Balmer (n=5) &1\\
1124.264 &6  &8 &?
         &\\
1136.003 &27 &18 &1136.532 &\ion{Ne}{v} 
         &$2s^2~2p^2$ $^3P_1$ -- $2s~2p^3$ $^5S_2$ &8 \\
1136.494 &4&3 &? \\
1136.775 &4&5 &? \\
1138.971 &4&2 &1139.494 &\ion{Fe}{ii}
         &$(^3F)4s$ b$^4F_{7/2}$ -- $5p$ $^4D_{5/2}$  ($*$) &7\\
1140.653 &8&6 &1141.172 &\ion{Fe}{ii} 
         &$4s$ b$^4F_{5/2}$ -- $5p$ $^4D_{5/2}$  ($*$) &7\\
1142.429  &9&7  &1142.956 &\ion{Fe}{iii} 
         &$3d^6$ $^3D_3$ -- $(^4G)4p$ $^3F_4$ ($*$) &7\\
1145.080 &62 &45 &1145.615 &\ion{Ne}{v} 
         &$2s^2~2p^2$ $^3P_2$ -- $2s~2p^3$ $^5S_2$ &8\\
1152.102 &8&7 &? \\
\enddata
\tablenotetext{a}{Wavelengths are accurate to around $\pm0.02$\,\AA.}
\tablenotetext{b}{Absolute fluxes accurate to 15~\% for strong lines,
        up to 40~\% for weakest lines.}
\tablenotetext{c}{Wavelength was derived by estimating the
center of the base of the line.}
\tablecomments{Wavelength references: (1) Appendix~A.3; (2) NIST atomic
database; (3) \citet{edlen83}; (4) \citet{boyce35}; (5) Appendix~A.2;
(6) \citet{kauf89}; (7) Kurucz; (8) Appendix~A.1.} 
\end{deluxetable}

\afterpage{\clearpage}

\begin{figure}[h]
\epsscale{0.7}
\plotone{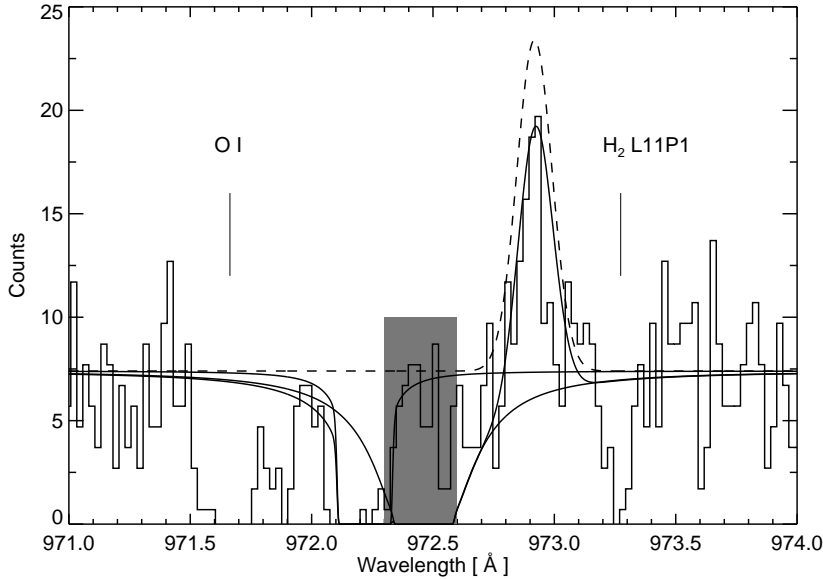}
\caption{SiC2A Spectrum from 2001~April showing the \ion{Ne}{vii}
\lam973 emission line. The spectrum has been binned by a factor 4. The
shaded region denotes Ly\gam\ airglow emission. The dashed and solid
lines demonstrate that the emission line escapes absorption by the
Ly\gam\ interstellar line (see text for details). Other interstellar
absorption lines due to \ion{O}{i} and H$_2$ are indicated.\label{ne7-model}}
\end{figure}

\section{Detection of Ne\,VII}

A weak line is detected at 972.917\,\AA\ which if corrected for
the AG Dra systemic velocity using the measurement for the \ion{S}{vi}
\lam944 line (Sect.~\ref{line-wid})
gives a rest wavelength of 973.353\,\AA. This matches
the 2s\,2p $^1P_1$ -- 2p$^2$ $^1D_2$ transition of
\ion{Ne}{vii} which has been detected in the laboratory by
\citet{edlen83} at 973.35\,\AA\ and in solar spectra by
\citet{feldman97}. The transition is  analogous to the
\ion{O}{v} \lam1371 line which is formed, at the low electron
temperatures found in the nebula, by recombination from \ion{O}{vi}
(Sect.~\ref{sect.nebtemp}). 
The identification of the \ion{Ne}{vii} line thus indirectly implies
the existence of \ion{Ne}{viii} in the AG Dra nebula.

The large radial velocity of AG Dra allows the \ion{Ne}{vii} line to
escape interstellar absorption by the H$_2$ L11P1 transition at
973.348\,\AA, yet it pushes the line closer to the interstellar
\ion{H}{i} Ly\gam\ absorption. Fig.~\ref{ne7-model} shows a comparison
of the SiC2A spectrum with a model spectrum, demonstrating that the
\ion{Ne}{vii} line is not significantly absorbed by Ly\gam. The model
shows a uniform continuum, and the \ion{Ne}{vii} \lam973.35 line,
shifted by $-133$~\kms. Two components were required to model the
Ly\gam\ absorption. The first is at a velocity of $-23$~\kms\ with a
column density of $5\times 10^{20}$~cm$^{-2}$ (a $b$ value of 10~\kms\
was chosen). This was estimated by
fitting the damping wings of the \ion{H}{i} Ly\be\
absorption line. The second component is at a velocity of $-108$~\kms\
and has a column density of $5\times 10^{19}$~cm$^{-2}$. This is
required to fit the absorption between 972.1 and 972.3\,\AA\ and give
the sharp edge at 972.1\,\AA.

No atomic data exist in the literature for interpreting the strength
of the \ion{Ne}{vii} line. We note that the flux is around a factor 10
weaker than the \ion{O}{v} \lam1371 line \citep{miko95}. The solar
photospheric O/Ne ratio is 5.6 \citep{grevesse98} and so the
\ion{Ne}{vii} flux is consistent, within an order of magnitude, if the
oxygen nebula conditions and atomic data are extended to neon.

The implied existence of \ion{Ne}{viii} in the AG Dra nebula is of
great importance as the ionization potential of Ne$^{+6}$ is 
207.3~eV making it the highest ionization species ever found in the
spectrum of AG Dra (e.g., the ionization potential of O$^{+5}$ is only
113.9~eV). It thus probes a much deeper part of the nebula.

\section{The O\,VI lines}

The two \ion{O}{vi} lines are the strongest lines in the \fuse\ spectrum
and both show two distinctive features: asymmetric profiles and
enhanced continuum at the bases of the lines. The $\lambda$1038 line
is affected by \ion{C}{ii}  $\lambda$1037.018 and H$_2$ L5R1
\lam1037.146 interstellar absorption on the short wavelength side of
the profile, which serves to make the $\lambda1032/\lambda1038$ flux
ratio larger than the optically thin ratio of 2.
The 
long wavelength side of \lam1038 is unaffected by the ISM and
closely matches the 
shape and position of the \lam1032 line in velocity space.

\subsection{Line profile}

\begin{figure}[h]
\epsscale{1.0}
\plottwo{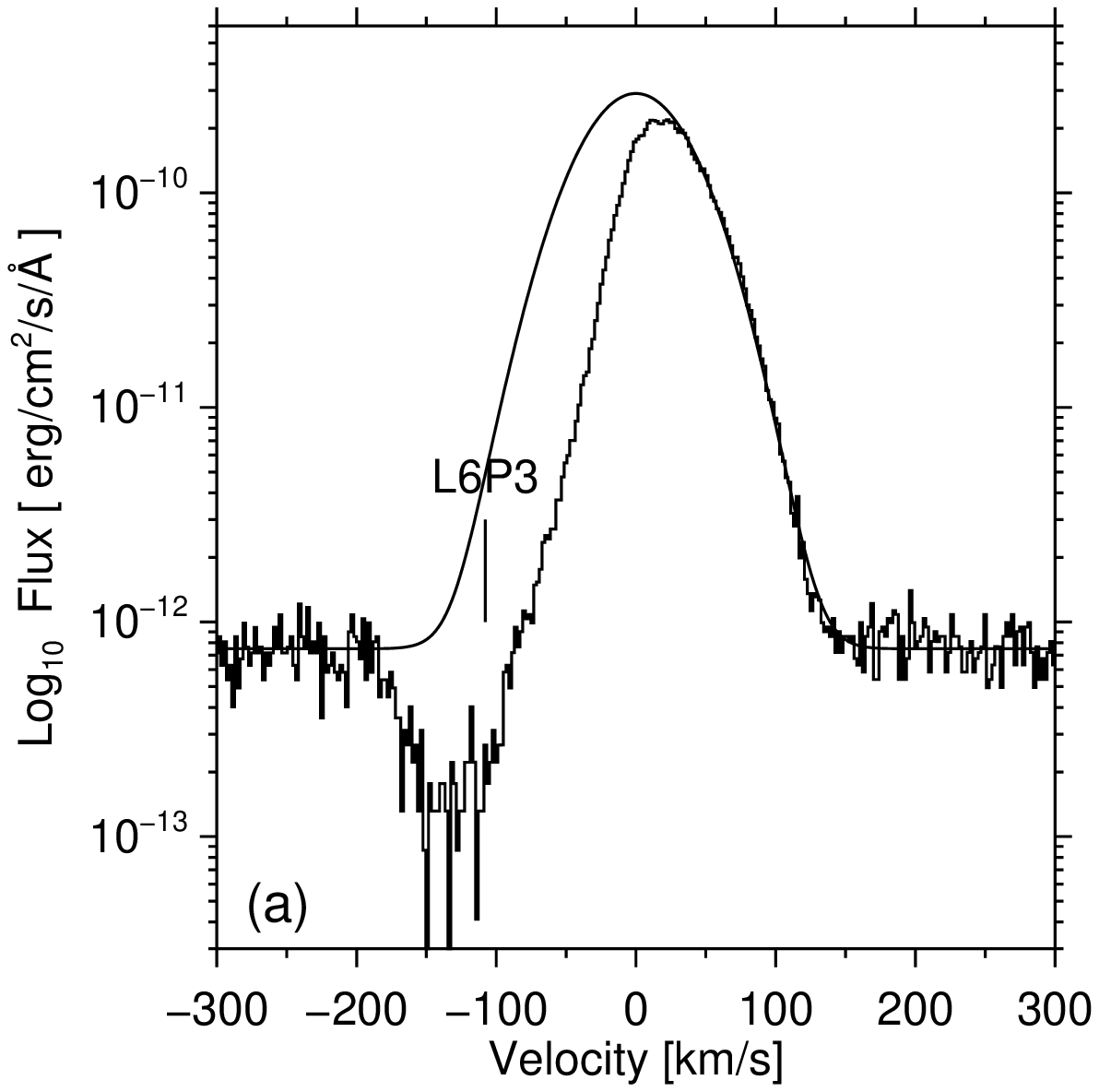}{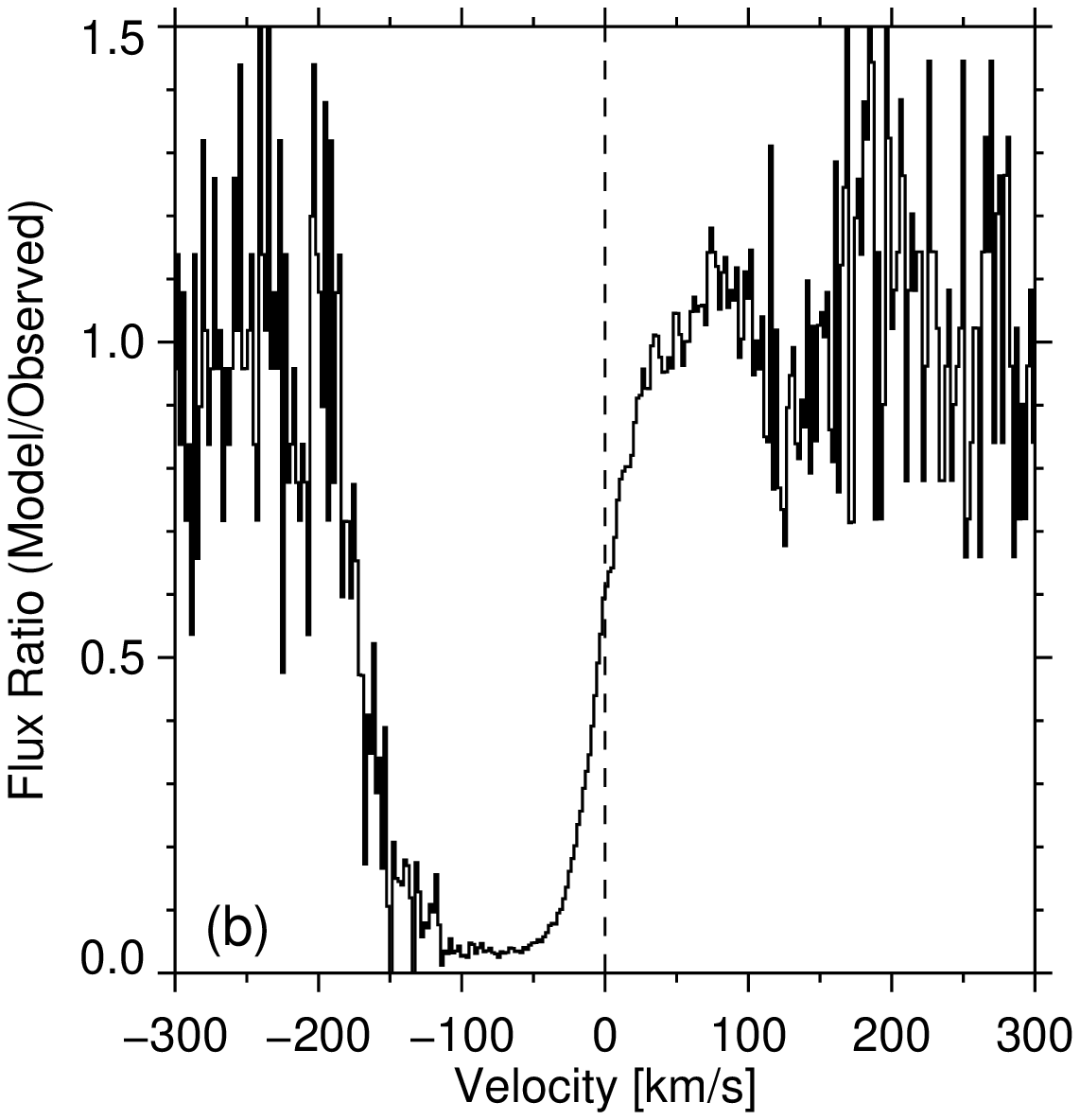}
\caption{(a) The LiF1A \ion{O}{vi} \lam1032 from the 2000~March
observation, plotted on a logarithmic scale, revealing the P Cygni
absorption on the short wavelength side of the emission line. The
expected location of the 
H$_2$ L6P3 interstellar absorption is indicated. A Gaussian profile
fitted to the continuum level and the long wavelength half of the
profile is overplotted. (b) The Gaussian profile shown in (a) is
divided through by the observed profile, revealing the full extent of
the absorption.\label{o6-gauss}}
\end{figure}

Fig~\ref{o6-gauss}(a) shows the $\lambda$1032 line plotted on a
logarithmic flux scale, which reveals a P Cygni-like profile for the
line. On this plot we have overplotted a Gaussian profile derived from
fitting only the continuum and the red side of the $\lambda$1032
profile. Also shown is the expected position of the L6P3 H$_2$ absorption
line. Clearly there is a possibility that it is the H$_2$ line giving
rise to a false P Cygni profile. Fig~\ref{o6-gauss}(b) shows the
actual spectrum divided by the computed Gaussian spectrum, revealing
the absorption to be $\approx 200$~\kms\ wide. This is much larger than
the width of other H$_2$ lines in the spectrum -- e.g., the nearby
L6R3 \lam1028.986 line has a FWHM of 30~\kms\ -- and so we attribute
the absorption 
to the \ion{O}{vi} profile. The fact that the blue wing of the
$\lambda$1032 line is being absorbed (in addition to the continuum)
shows that the \ion{O}{vi} emitting region 
is an expanding, optically thick medium.

A P Cygni profile in the \ion{N}{v} $\lambda$1238 line seen in
spectra
from
\emph{IUE} was first reported by \citet{viotti83}, and discussed later
by \citet{viotti84}, \citet{lutz87} and
\citet{kafatos93}. In all these cases the P Cygni profile was only
found during the outburst phase of AG Dra. This may simply be due to
the low signal-to-noise in the continuum during quiescent periods,
however, and \citet{viotti84} note that the $\lambda$1238 line
``$\ldots$is always characterized by a sharper blue wing$\ldots$'', an
asymmetry consistent with the profile found for \ion{O}{vi}
$\lambda$1032 here. The significance of the \ion{O}{6} profile in
relation to the structure of the nebula is discussed in
Sect.~\ref{sect.discussion}.

\subsection{Electron scattering wings of the O\,VI
  doublet}\label{sect.o6-wings} 

\begin{figure}[t]
\epsscale{1.0}
\plotone{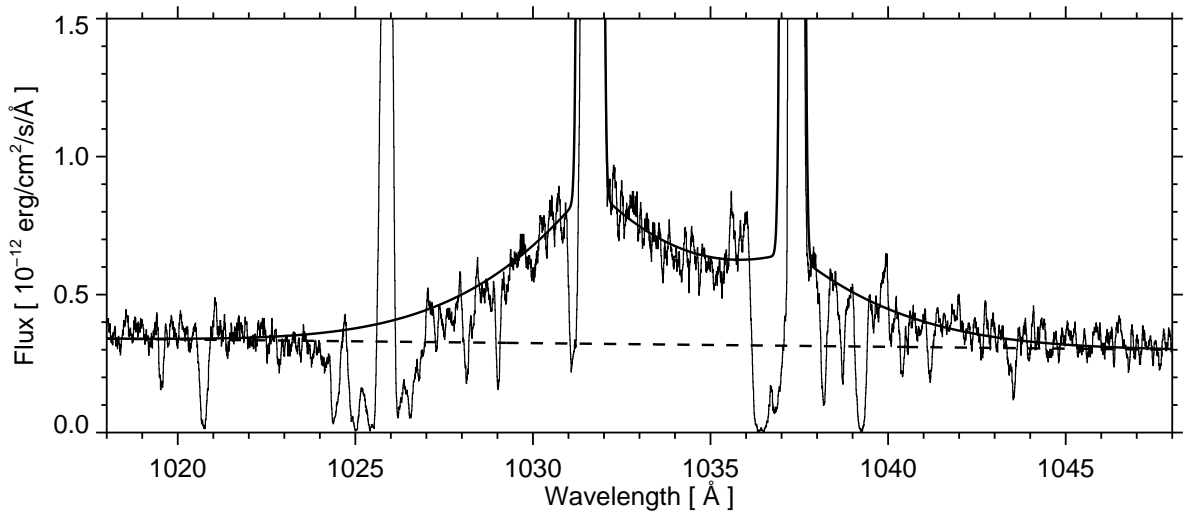}
\caption{The LiF1A spectrum from 2000~March showing the broad wings of
the \ion{O}{vi} \lam1032 and \lam1038 lines. Overplotted is a model
for the electron scattering wings assuming an electron temperature of
30,000~K and an optical depth of 0.04.}
\label{o6-broad-wings}
\end{figure}

The broad wings to the \ion{O}{vi} profiles were measured previously by the
ORFEUS echelle spectrograph \citep{schmid99}. They can be compared to
the wings measured for the \ion{He}{ii} $\lambda$1640 line by
\emph{IUE} \citep{viotti83}. Possible instrumental causes of the wings
(e.g., grating scattering or `bleeding' of the detector electronics)
can be ruled out by the presence of sharp interstellar absorption
lines in the wings and the fact that the absorption in the \ion{C}{ii}
interstellar lines goes to zero. \citet{schmid99} suggested that the
wings are caused by Thomson scattering of \ion{O}{vi} photons from
nebular electrons, and equated the width of the wings to an electron
temperature of 30\,000~K. Using the expression of \citet{bernat78} for
electron scattering in a layer outside of the line formation region,
together with the Thomson scattering redistribution function of
\citet{mihalas70} we find an optical depth of 0.04 and an electron
temperature of 30\,000~K fits the \ion{O}{vi} $\lambda\lambda$1032,
1038 line wings reasonably well (Fig.~\ref{o6-broad-wings}). These
values of the optical 
depth and electron temperature can be used to estimate a size and
density of the 
\ion{O}{vi} region. Since $\tau\approx \sigma_{\rm T}N_e r$, where
$\tau$ is the optical depth, $\sigma_{\rm T}$ the Thomson
cross-section, $N_e$ the electron density and $r$ the radius of the
(assumed) spherical \ion{O}{vi} emitting region, and the flux
in the \ion{O}{vi} $\lambda$1032 line is related to $N_e$ and $r$ as
(assuming the line is optically thin)

\begin{equation}\label{eq.f-vol}
4\pi d^2 F=0.83 {hc\over \lambda} Ab({\rm O}) {n_j\over N_e} A_{ji} N_e^2 {4\pi
r^3\over 3}
\end{equation}

\noindent where $d$ is the distance to AG Dra, $F$ the line flux, the
factor 0.83 represents the ratio of protons to electrons in a fully
ionized plasma, $h$ the Planck
constant, $c$ the speed of light, $\lambda$ the wavelength, $Ab({\rm
X})$ the abundance of element X relative to hydrogen, $n_j$ the population of the emitting
level of the $\lambda$1032 transition relative to the ion population,
and $A_{ji}$ is the radiative decay rate for the transition. The O/H
abundance is obtained from \citet{smith96}, who measured the
photospheric abundance in the giant, while the value of $n_j$ was
obtained from the model of \ion{O}{vi} in the CHIANTI database
\citep{young03, dere97}, assuming an electron temperature of
30\,000~K, and an electron density of $10^{10}$~cm$^{-3}$ (note that
$n_j/N_{\rm e}$ is independent of $N_{\rm e}$ for the \lam1032 line,
and so this choice does not affect the derived result). The $A_{ji}$
value was also obtained from CHIANTI.

The two  relations then imply
\begin{eqnarray}
N_e r &\approx& 7.52 \times 10^{22} ~{\rm cm}^{-2}\\
N_e^2 r^3 &\approx& 4.15 \times 10^{57} ~{\rm cm}^{-3}.
\end{eqnarray}
Thus one has $r\approx 17.3\times 10^{11}$~cm ($=11~R_\odot$) and
$N_e\approx 1.0\times 10^{11}$~cm$^{-3}$. Given the uncertainties in
the various parameters, we estimate that these quantities are accurate
to no better than a factor of 5. The density is an order of magnitude
higher than that derived from the \ion{O}{iv} density diagnostic
(App.~\ref{app.o4}), which one may expect if the \ion{O}{vi} is formed
in a deeper layer of the nebula.

This same calculation was performed by \citet{viotti83} when
interpreting the broad wings of the \ion{He}{ii} $\lambda$1640 line
observed by \emph{IUE} during an active phase of the system. They
ruled out the possibility of electron scattering on account of the
large electron density ($5.0\times 10^{12}$~cm$^{-3}$) they derived
which was inconsistent with the density of $\sim 10^{10}$cm$^{-3}$
derived from 
the \emph{IUE} \ion{O}{iv} density diagnostic (see also
Appendix~B). We note that using the 2500~pc distance to AG Dra instead
of their 700~pc distance increases their \ion{He}{iii} region radius
to $7.7\times 10^{11}$~cm and decreases their density to $3.9\times
10^{11}$~cm$^{-3}$, in reasonable agreement with the \ion{O}{vi}
numbers. The optical depth of the electron layer is much
greater from the \ion{He}{ii} line than from the \ion{O}{vi} lines,
although one may expect the opposite case as \ion{O}{vi} will be
formed much deeper inside the nebula. However, \citet{miko95} found
the AG Dra nebula to be larger during outburst, when the
\citet{viotti83} measurement was made, which would thus lead to a larger
optical depth.

Another possibility for the broad wings is that they represent the
thermal emission of a fast wind with speeds 1000--2000~\kms. Such a
scenario is presented in the AG Peg observation of \citet{nussb95},
however in this case the broad wings are accompanied by a P Cygni
profile in the \ion{N}{v} $\lambda$1238 line that extends out to
velocities $\approx 1150$~\kms. For AG Dra, the P Cygni absorption
reaches only $\approx 200$~\kms, and so it is unlikely that the broad
wings are due to a fast wind. In addition AG Peg has a history of high
velocity Wolf-Rayet features \citep{kenyon93} while AG Dra has no such
history.

\section{Ne\,VI lines}\label{sect.ne6}

The five \ion{Ne}{vi} $2s^22p$ $^2P_J$ -- $2s2p^2$ $^4P_{J^\prime}$ intercombination lines are
prominent in the AG Dra spectrum 
between 992 and 1010\,\AA. The \lam992 line is the weakest of the
multiplet and found in the wing of \ion{He}{ii} \lam992.363. Due
to the large radial velocity of AG Dra the \ion{Ne}{vi}
\lam1005.696 line is almost exactly coincident with the L8P2 H$_2$
1005.397\,\AA\ interstellar absorption line, leading to a
double-peaked profile for the emission 
line. 
The \ion{Ne}{vi} \lam1010.247 line is truncated on the short
wavelength side of the profile by the H$_2$ W0Q1 \lam1010.132
absorption. The \lam997 and \lam999 lines are free from interstellar
absorption and blending, and their ratio potentially allows a
constraint on the nebular electron density to be 
made \citep{espey96}.

The model we use for the \ion{Ne}{vi} ion employs the electron
excitation and radiative decay rates from v.4 of the CHIANTI atomic
database \citep{young03, dere97}. Proton excitation
rates are also included, but are negligible for the low temperatures
found in nebular 
conditions. An important process to account for is photon excitation
by the white dwarf radiation field. A star of effective temperature
1--$2\times 10^5$~K emits predominantly in the extreme ultraviolet,
between wavelengths 100--600\,\AA. The strong $2s^22p$--$2s2p^2$
resonance transitions of \ion{Ne}{vi} have energies between
400--600\,\AA\ and so will 
be strongly pumped by such a radiation field at distances close to
the star.

\begin{figure}
\epsscale{0.7}
\plotone{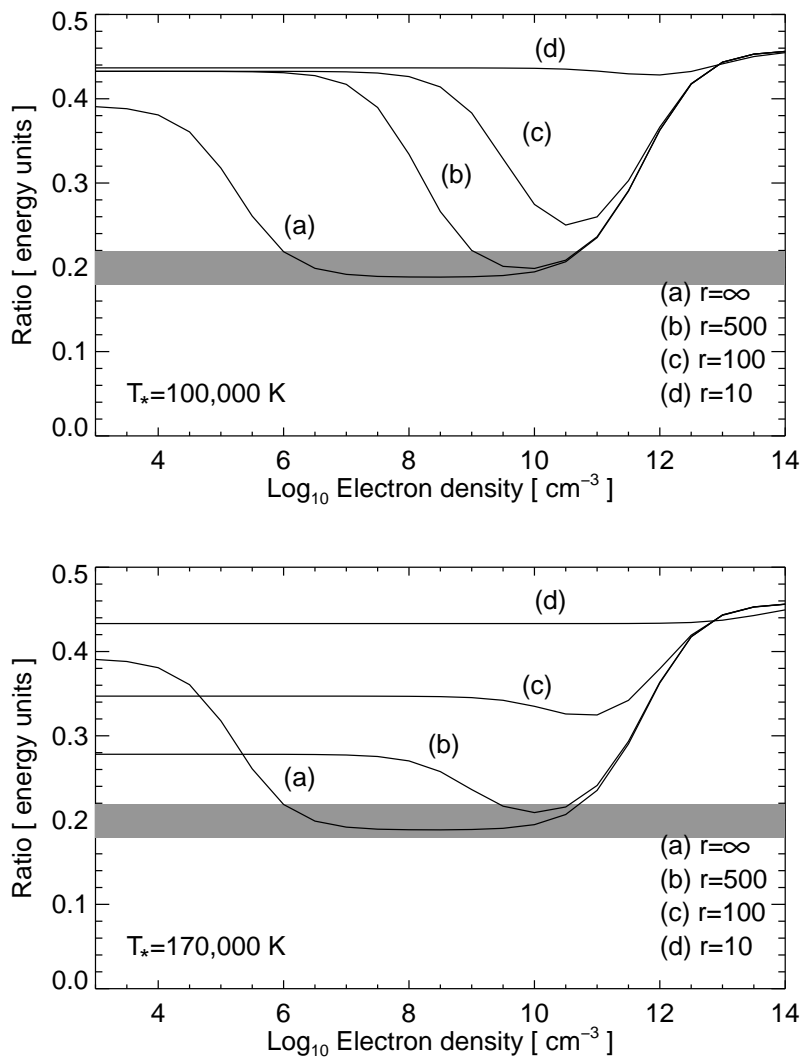}
\caption{Predicted variation of the \ion{Ne}{vi} \lam997.14/\lam999.26
emission line ratio in the presence of 100,000~K (top panel) and
170,000~K (bottom panel)
blackbody radiation fields. Four cases are shown, corresponding to the
\ion{Ne}{vi} ions being different distances ($r$, in white dwarf radii
units) above the surface of the white dwarf. The grey band denotes the
measured \lam997/\lam999 ratio from the 2000~March \fuse\ spectra.
Atomic data
are from the CHIANTI database.\label{ne6-ratio}}
\end{figure}

Photon excitation can be accounted for in
v.4 of CHIANTI through specifying a black body radiation field of
temperature $T_*$, and the distance from the star's centre, $r$,
measured in stellar radii units \citep{young03}. For the
AG Dra model, we consider two temperatures: the value of 100,000~K
derived by \citet{miko95} from fits to the UV continuum measured by
\iue; and the value of 170,000~K derived by \citet{greiner97} from
X-ray data from ROSAT. An electron temperature of 25,000~K is
assumed. Figs.~\ref{ne6-ratio}a,b show the effects on the
\ion{Ne}{vi} \lam997/\lam999 ratio for each temperature. The four
curves in each plot correspond to different values of $r$. The shaded
region on each plot denotes the measured 
ratio, with 1$\sigma$ error bars.
It is clear that the observed \lam997/\lam999 ratio is
incompatible with the \ion{Ne}{vi} ions being located at
$\lesssim$500 white dwarf radii ($R_{\rm wd}$) from the white
dwarf, and thus photoexcitation has a weak effect on the level
populations. For a white dwarf radius of 
0.1~$R_\odot$ \citep{miko95}, this 
places the \ion{Ne}{vi} ions at least 50~$R_\odot$ from the white
dwarf and a significant way towards the giant at around
300~$R_\odot$ \citep{miko95}. The ionization potential of Ne$^{+4}$ is
126.2~eV, which compares with 113.9~eV for O$^{+4}$, thus implying that the
\ion{O}{vi} emission is from regions even further from the white
dwarf. One can 
thus rule out the suggestion \citep[e.g.][]{viotti83,miko95} that
\ion{He}{ii} and \ion{N}{v} emission is from very close to the white
dwarf.

Our \ion{Ne}{vi} model can be affected if the actual white dwarf
spectrum deviates significantly from a black body. In particular, the
very transitions that are pumped in the \ion{Ne}{vi} ion are likely to
appear as absorption features in a hot white dwarf atmosphere. A
Kurucz spectrum calculated for an effective temperature of 100,000~K
and $\log\,g=8.0$, with solar abundances, shows \ion{Ne}{vi}
absorption lines whose depth at line center is around 30-60\% of the
continuum level. Thus photoexcitation may be overestimated by a
factor of 2--3 in the black body model. We thus suggest that the
\lam997/\lam999 ratio constrains the \ion{Ne}{vi} ions to be formed at
a distance of at least 300 white dwarf radii.

In the case of a negligible radiation field, the electron density is
restricted to the range $5.5\le \log\,N_{\rm e}\le 10.5$. Increasing
the radiation dilution factor narrows the allowed densities towards
the higher end 
of this range, with the case of $r=500$ implying densities of
$\sim 10^{10}$~cm$^{-3}$. This is consistent with densities derived from
the intercombination transitions of \ion{O}{iv}, measured with \iue\
\citet{viotti83}. Fluxes obtained with \iue\ are
reassessed in Appendix~B with the most recent atomic data from
CHIANTI. A density 
of 10$^{10}$~cm$^{-3}$ is derived.

\section{The He\,II recombination lines}

The Balmer series of \ion{He}{ii} (transitions n$\rightarrow$2, n$\ge$3)
provide many emission lines in the AG Dra spectrum, from the n=5 line
at 1085\,\AA\ to the n=20 line at 920\,\AA. The even members of the
sequence lie close in wavelength to the \ion{H}{i} Lyman series, at
around 0.3\,\AA\ shortward of the \ion{H}{i} lines.
This, combined with the $\sim 100$~\kms\ blueshift of the
stellar lines compared to the interstellar absorption, is sufficient to
make all lines from n=6 and above observable. Coincident blends with
other interstellar absorption species prevent the strongest n=6 and n=8 lines
being observed, but the n=10, 12 and 14 lines are all visible. These
three lines are factors of 2--3 weaker than expected from the odd
members of Balmer sequence and are blueshifted relative to these lines
by $\approx$5--10~\kms\ (Sect.~\ref{line-wid}). This behaviour is puzzling. 
The reduced fluxes are not due to an intrinsic property of the
\ion{He}{ii} ions as recombination does not favor odd over even
members of the sequence. Any absorption (either stellar or
interstellar) must occur on the long wavelength sides of the lines to
give rise to the observed blueshifts. The natural candidate is
absorption through the damping wings of the hydrogen lyman lines. 
However, both the \lam937 and \lam949 lines show continuum on their
long wavelength sides which is at the same level as the continuum in
the rest of the spectrum, ruling out absorption by the \ion{H}{i}
lines.

\begin{figure}
\epsscale{1.0}
\plotone{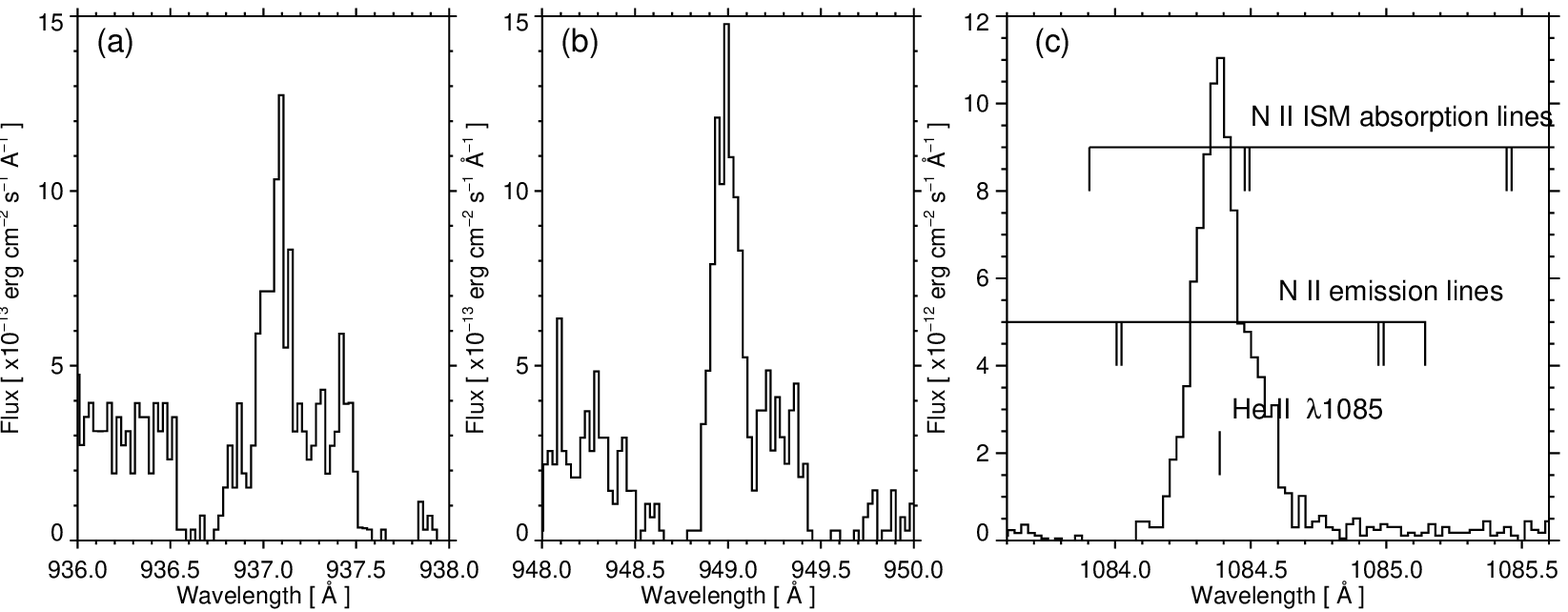}
\caption{Plots showing the \ion{He}{ii} \lam937, \lam949 and \lam1085
emission lines (Balmer n=12, 10 and 5, respectively). Figures (a) and
(b) show continuum emission on the long wavelength side of the
emission lines, demonstrating that the reduced fluxes in the lines are
not due to interstellar hydrogen absorption. Figure (c) shows the
expected positions of  \ion{N}{ii} transitions, both in the cases of
them being due to interstellar absorption or being in emission from
the star.\label{he2-spec}}
\end{figure}

The strongest of the \ion{He}{ii} lines at \lam1085 is affected by
interstellar absorption in the \ion{N}{ii} \lam\lam1084.566, 1084.584
lines which result in an asymmetric appearance to the \lam1085
profile. We estimate that the absorption removes around 30\% of the
\ion{He}{ii} flux. 

The \ion{He}{ii} Balmer lines can be used to place constraints on the
density and temperature of the nebula. \citet{storey95} provide
emissivities for all of the lines up to n=50, calculated assuming
Cases A and B of \citet{baker} over a wide range of
temperatures and densities. Case A requires the \ion{He}{ii} Ly\al\
line to be optically thin, which is not applicable for the AG Dra
nebula on account of the high densities.
We consider the lines \lam958, \lam942 and \lam927 (n=9, 11, and 15,
respectively) which are unblended and not contaminated by interstellar
absorption. The closeness of the three lines in wavelength minimizes
any distortions in the line ratios due to dust extinction.

\begin{figure}[h]
\epsscale{0.7}
\plotone{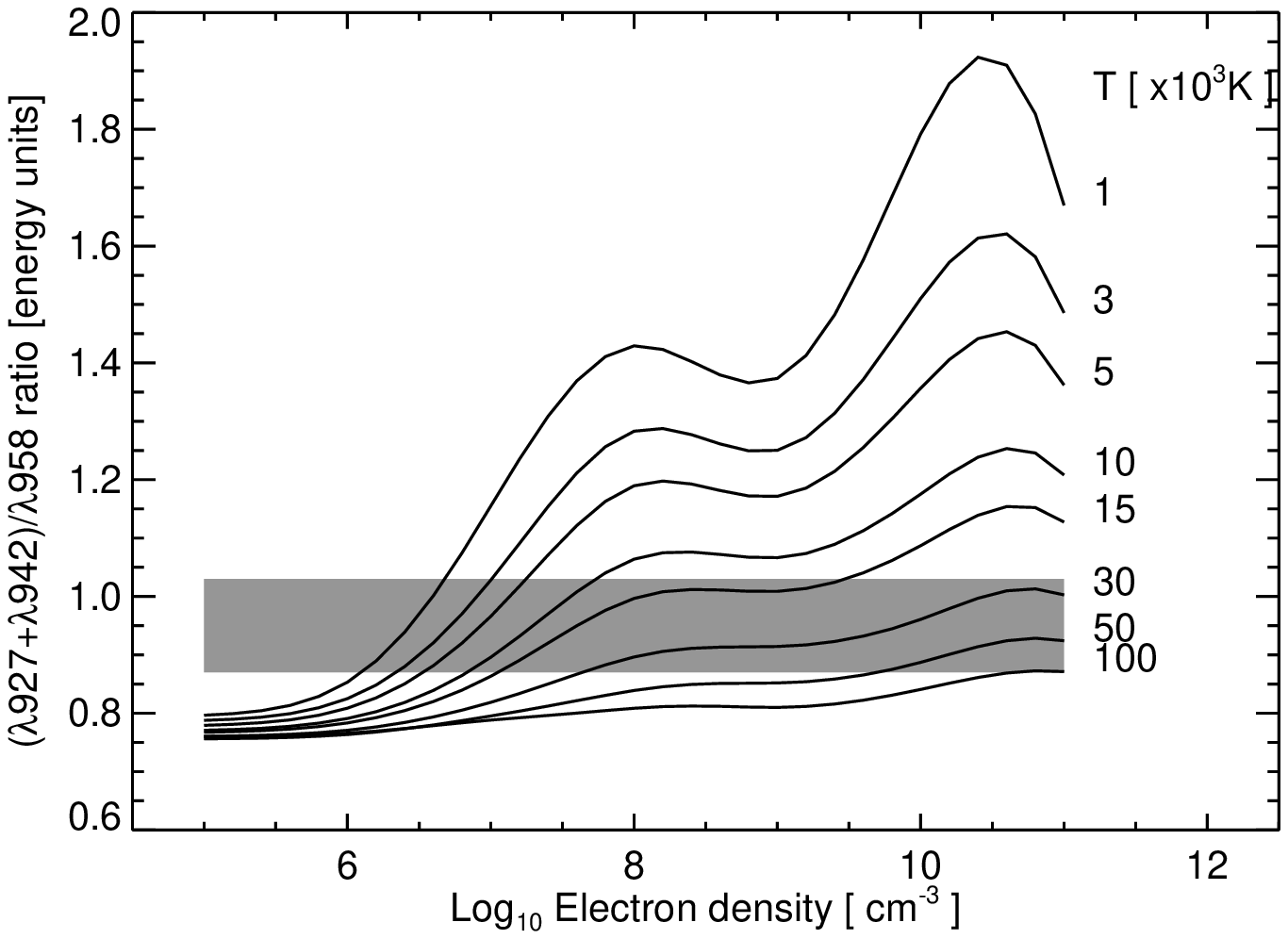}
\caption{The lines show the predicted variation of the \ion{He}{ii}
(\lam927+\lam942)/\lam958 ratio with density for a range of electron
temperatures, derived from the calculations of \citet{storey95}. The
shaded region shows the $\pm1\sigma$ errors on the observed
ratio.\label{he2-958}} 
\end{figure}

Fig.~\ref{he2-958} plots the theoretical variation of the
(\lam927+\lam942)/\lam958 ratio with density for a wide range of
temperatures. Also plotted is the observed ratio derived from
combining the SiC1B and SiC2A spectra from the 2001 April
observation, and which has a value of $0.95\pm 0.08$. It is clear that
low nebula temperatures ($T\le 10^4$~K) are ruled out unless the
density is also low ($N_{\rm e}\lesssim 10^8$~cm$^{-3}$), while for hotter
nebula temperatures low densities are ruled out. Other plasma
diagnostics (Sects.~\ref{sect.nebtemp} and Appendix~B) suggest a
density of $10^{10}$~cm$^{-3}$ and a temperature of $\approx 2\times
10^4$~K and these values are in reasonable agreement with the
\ion{He}{ii} lines.

The \ion{He}{ii} lines can also be useful in determining the
extinction towards AG Dra. We consider the \lam958, \lam1085 and
\lam1640 lines (n=9, n=5 and n=3, respectively), the latter of which
was observed many times with \iue\ and was found to have a flux
between 
 $2\times 10^{-11}$ and $8\times
10^{-11}$~\ecs\ during the quiescent phase of AG Dra
\citep{griestra99}. Assuming no extinction and nebula conditions of $T=2\times 10^4$~K
and $N_{\rm e}=10^{10}$~cm$^{-3}$, \citet{storey95} predict
\lam1640/\lam958 and \lam1640/\lam1085 ratios of 24.2 and 5.49,
respectively. Including the full range of variability of the \lam1640
line gives upper and lower limits of 143 and 36 for \lam1640/\lam958,
and 24 and 6.0 for  \lam1640/\lam1085 (adjusting the \lam1085 flux for
interstellar absorption). These values show
the observed \lam958 and \lam1085 lines are weaker than expected from
the  \lam1640 line, consistent with a non-zero $E_{B-V}$. Assuming a
more precise \lam1640 flux of $3.5\times 10^{-11}$ based on 
the orbital phase of the 2001~April observation, and the extinction
curve of \citet{fitz99} allows us to estimate an $E(B-V)$ value of
0.10 for AG Dra.
This value is significantly higher than that
found by \citet{miko95}, but we note that the \citet{fitz99}
extinction curve is an extrapolation to \fuse\ wavelengths based on
observations made above $\approx$1200\,\AA\ and so may be in error.

\section{The Nebula Temperature}\label{sect.nebtemp}

Methods for determining the electron temperature in the nebulae of
symbiotic stars from emission line ratios are discussed in Sect.~3 of
\citet{nussb87}, with reference to \iue\ spectra. Temperatures for the
nebulae were typically found to be $\lessapprox$20,000~K, indicating that
the radiation field from the hot component of the symbiotic dominates
the ionization balance of the plasma, rather than electron collisions.
The \fuse\
wavelength range opens up further diagnostic possibilities, and the
use of \ion{O}{vi} and \ion{C}{iii} lines to constrain the plasma
temperature are discussed below.

\subsection{O\,VI, O\,V temperature diagnostic}\label{sect.o6-temp}

In a highly-ionized photoionized plasma such as the AG Dra nebula, it
is possible to use the \ion{O}{vi} \lam1032 and \ion{O}{v} \lam1371
emission lines to determine the nebula temperature. 
For electron temperatures $\sim 10^4$~K the $2s2p$ $^1P_1$ -- $2p^2$ $^1D_2$
transition of beryllium-like ions (e.g., \ion{O}{v} \lam1371) is
predominantly formed through 
dielectronic recombination from the lithium-like ion (e.g.,
\ion{O}{vi}) into the
$^1D_2$ level. The emissivities of such \emph{recombination lines}
vary slowly with 
temperature. In contrast, for emission lines excited directly
through electron collisions, the emissivity falls sharply with
temperature on account of the Maxwellian $\exp(-\deltE/kT)$ term in the
expression for the excitation rate coefficient. 
For temperatures $\gtrsim10^4$~K, the \lam1032 line is sufficiently
strong that it remains electron-excited rather than
recombination-excited, and so the \ion{O}{vi} \lam1032/\ion{O}{v}
\lam1371 ratio is an excellent temperature diagnostic, independent of
the relative concentrations of the two ions.

\begin{figure}[h]
\epsscale{0.5}
\plotone{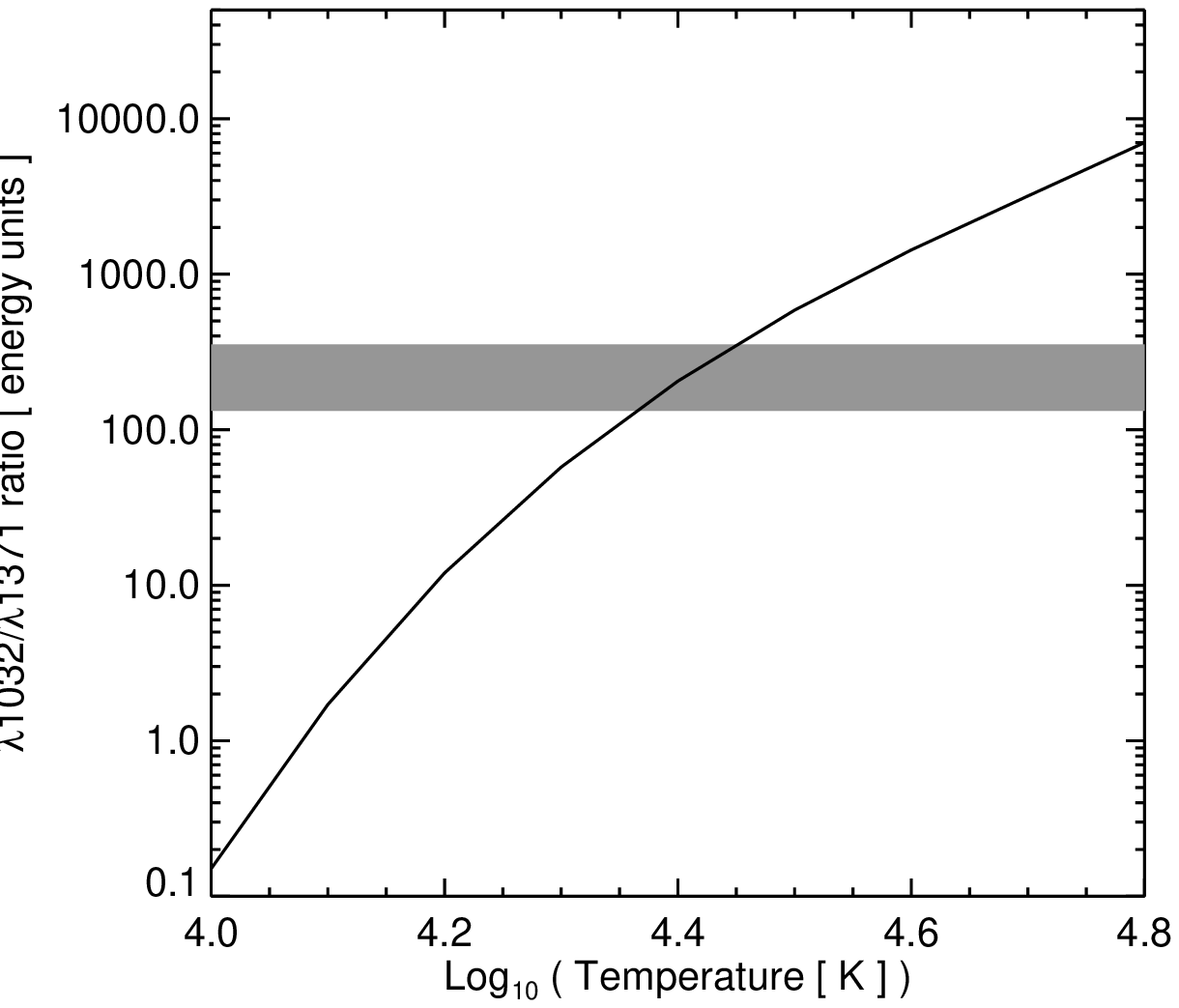}
\caption{Theoretical variation of the \ion{O}{vi} \lam1032/\ion{O}{v}
\lam1371 ratio. The dashed lines indicate upper and lower limits to
the observed ratio based on the present \fuse\ observations of the
\lam1032 line and \iue\ observations of the \lam1371
line.\label{o5-o6}} 
\end{figure}

Recombination data suitable for predicting the strength of the
\ion{O}{v} \lam1371 line are from \citet{nussb84}, while data for the
\lam1032 line 
are from v.4 of the CHIANTI database \citep{young03}. The
\lam1032/\lam1371 ratio varies
by several orders of magnitude over a small temperature 
range (Fig.~\ref{o5-o6}), providing a tight constraint to the electron
temperature. Measurements of the \lam1371 line at different orbital phases
of AG Dra were presented by \citet{miko95}. Using these values
together with the \ion{O}{vi} \lam1032 flux from \fuse\ gives the ratio
indicated in Fig.~\ref{o5-o6}, yielding a temperature of $\approx
25,000$\,K, in good agreement with the temperature of 30,000~K 
derived from the electron scattering wings on \ion{O}{vi} line
profiles.

\subsection{C\,III}

\ion{C}{iii} gives rise to two dominant spectral features in the \fuse\
waveband: the $2s$ $^1S$ -- $2p$ $^1P$ transition at 977.020\,\AA\ and the
set of six $2s2p$ $^3P_J$ -- $2p^2$ $^3P_{J^\prime}$ transitions
between 1174 and 
1176\,\AA. These lines together with the \lam1909 line observed by
\iue\ potentially allow both the temperature and density of the plasma
to be estimated.

\begin{figure}[h]
\epsscale{0.9}
\plotone{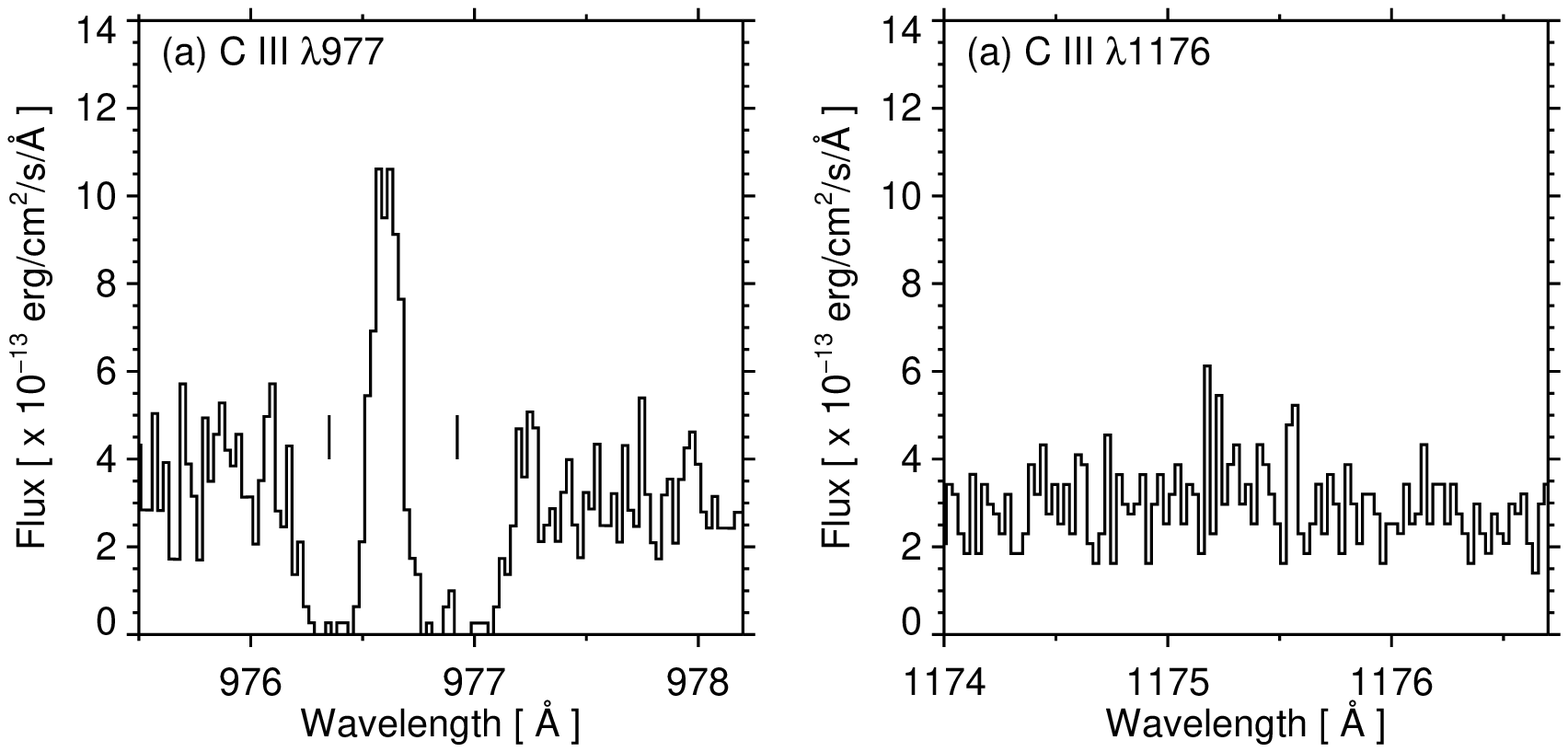}
\caption{SiC2A spectrum from 2001~April, showing the \ion{C}{iii}
\lam977 emission line. The absorption features on each side of the
emission line are due to interstellar absorption by \ion{O}{i}
\lam976.448 and \ion{C}{iii} \lam977.020.\label{fig.c3-plot}}
\end{figure}

The \lam977 profile from the 2001 April observation is
displayed in Fig.~\ref{fig.c3-plot} and shows the emission line to lie
within a deep absorption trough in the white dwarf continuum. The
absorption on the long wavelength side of the profile is due to
\ion{C}{iii} in the interstellar medium -- the AG Dra emission line is
saved on account of the high radial velocity of the system. To the
short wavelength side is the \ion{O}{i} \lam976.448 interstellar
absorption line. The centroid of the emission line is redshifted by
$\approx 5$--15~\kms\ relative to other species indicating that it is
partially absorbed by the ISM components.

The \lam1176 feature is not detected (Fig.~\ref{fig.c3-plot}) and we
estimate an upper limit to 
the flux of $4\times 10^{-14}$~\ecs. The \lam1909 was detected by
\iue, and \citet{miko95} give quiescent fluxes of
1--4$\times$$10^{-13}$~\ecs.

\begin{figure}[h]
\epsscale{0.8}
\plotone{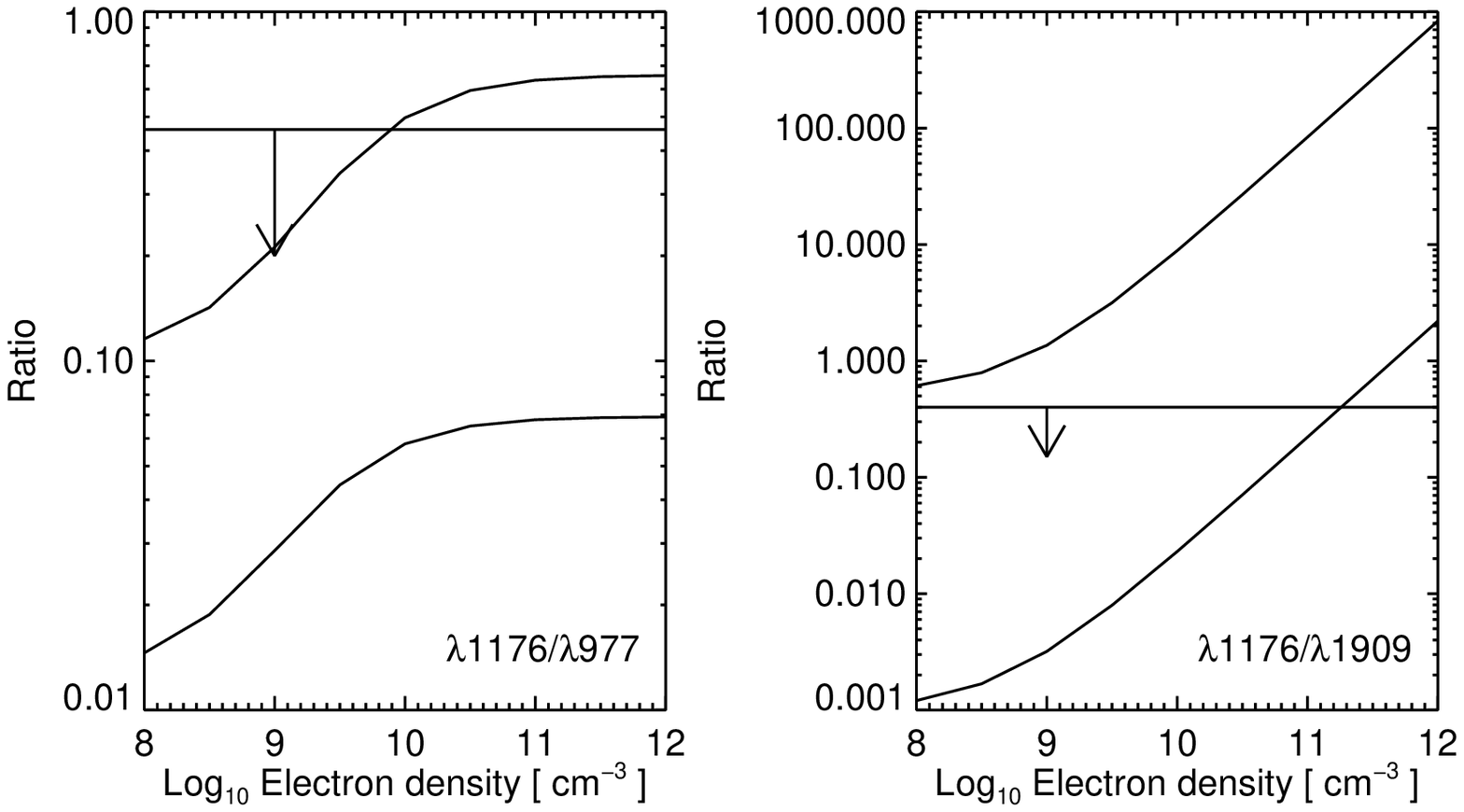}
\caption{Plots of the \ion{C}{iii} \lam1176/\lam977 and
\lam1176/\lam1909 ratios as a function of 
density at temperatures of 25\,000~K (upper curve) and 80\,000~K
(lower curve). The horizontal lines show upper limits derived from \fuse\
and \iue\ spectra. Calculated using the CHIANTI atomic database.
\label{c3-chianti}}
\end{figure}

Fig.~\ref{c3-chianti} shows the theoretical variation of the
\lam1176/\lam977 and \lam1176/\lam1909 ratios, derived from the
CHIANTI database. In each case the ratios are plotted at two
temperatures: 80,000~K and 25,000~K, the former corresponding to the
temperature of maximum abundance of \ion{C}{iii} assuming an
electron-ionized plasma, and the latter the temperature derived in
Sect.~\ref{sect.o6-temp} for the \ion{O}{vi} region of the AG Dra
nebula.

The measured \lam977 flux is a lower limit as the extent of
interstellar absorption is not known. The resulting \lam1176/\lam977 ratio
can thus not discriminate the temperature of the plasma.  
Using the lowest of the measured \lam1909 fluxes, the upper limit to
the \lam1176/\lam1909 ratio is overplotted in
Fig.~\ref{c3-chianti}(b). Clearly the ratio is incompatible with a
temperature of 80,000~K. A deeper exposure of the \lam1176 line
together with a simultaneous and more accurate measurement of the
\lam1909 line are required 
to improve the constraints on nebula temperature.

\section{Fluoresced and unidentified lines}\label{sect.fluor}

There are several emission lines in the AG Dra spectrum for which no
obvious identifications could be found based on solar spectra. These
lines are listed in Table~\ref{unid-corr}, together with their
wavelengths corrected by a velocity shift of $+136$~\kms, which is the
velocity of the \ion{Ne}{v} \lam1136 line. The corrected wavelengths
should be within $\pm$10~\kms\ of the laboratory wavelengths of the
transitions.

\begin{deluxetable}{ll}
\tablecaption{Wavelengths for unidentified lines.\label{unid-corr}}
\tablehead{Measured &Corrected  \\
          wavelengths &wavelengths \\
          (\AA) &(\AA) 
           }
\tablewidth{2.2in}
\startdata
1065.752 &1066.141 \\
1136.494 &1137.010 \\
1136.775 &1137.291 \\
1138.971 &1139.488 \\
1140.653 &1141.171 \\
1142.429 &1142.948 \\
1152.102 &1152.625 \\
\enddata
\end{deluxetable}

As many unidentified emission lines exist in the UV spectrum of the
symbiotic nova RR Telescopii that were later identified as
\ion{Fe}{ii} transitions fluoresced by the strong nebular lines
\citep{johan88}, we  searched for \ion{Fe}{ii} lines in the \fuse\
bandpass that could be similarly excited.  We initially searched for
the lines identified by \citet{harper01} who found many \ion{Fe}{ii}
lines fluoresced by \ion{H}{i} Ly\al\ in the \fuse\ spectrum of the
hybrid supergiant $\alpha$ TrA. None of the unidentified \fuse\ lines
matched the Harper et al.\ list. In particular, the 1133.70\,\AA\ and
1139.02\,\AA\ lines, pumped via a line only 0.182\,\AA\ from Ly\al\
line center, are not present implying either a narrow Ly\al\ line of FWHM
$\lessapprox$ 0.36\,\AA, or that the Ly\al\ line is formed far from
the surface of the giant.

In order to find other fluorescence lines  in
the \fuse\ waveband a simple model was developed to predict spectra
arising through radiative pumping of \ion{Fe}{ii} and \ion{Fe}{iii}
by a Gaussian shaped 
emission line at a specific wavelength. The model calculates the
quantity

\begin{equation}
\phi(\lambda)  {1 \over \lambda_{kj}} N_i  A_{kj} 
        \left[ A_{ki} {\omega_k \over \omega_i} \lambda_{ki}^3 \right]
\end{equation}

\noindent where $i$, $j$ and $k$ are indices for the \ion{Fe}{ii} levels such
that the transition 
$i\rightarrow k$ is the transition pumped by the emission line and
$k\rightarrow j$ is the subsequent radiative decay; 
$\phi(\lambda)$ is the intensity of the fluorescing emission line profile at
wavelength $\lambda$; $N_i$ is the population of level $i$  relative
to the ion population; 
$A_{kj}$ the radiative decay rate for transition $k\rightarrow j$;
 and $\omega_i$ is the statistical weight of
level $i$. The model allows the centroid and width of the fluorescing
emission line to be varied.
The $N_i$ are distributed according to the Boltzmann distribution, and
the 
energy level and radiative data for \ion{Fe}{ii} and \ion{Fe}{iii} are
from the compilations of R.L.~Kurucz, available at
http://cfaku5.harvard.edu/.

We considered  pumping of both \ion{Fe}{ii} and \ion{Fe}{iii} by the
\ion{O}{vi} \lam\lam1032, 1038 doublet, and Table~\ref{fluor-flux}
shows the model predictions. A \lam1038/\lam1032 ratio of 1:2 was
assumed, and the full width at half maxima of the \ion{O}{vi} lines
was taken as 0.30\,\AA. For both \ion{Fe}{ii} and \ion{Fe}{iii} there
is only one significant transition 
that is pumped by the \ion{O}{vi} lines; the resulting fluoresced
lines thus have relative fluxes determined purely by the radiative
decay rates for the transitions. 

The \ion{Fe}{ii} lines fluoresced by \lam1032 are all pumped through
the same \ion{Fe}{ii} transition. The \lam1776, \lam1881 and \lam1884
lines have been identified by \citet{johan88} in IUE spectra of RR Tel
and have also been measured in IUE spectra of AG Dra by
\citet{viotti83} although the authors did not identify the
transitions. The strongest of the predicted lines below 1200\,\AA\ is
at 1141.172\,\AA\ which provides an excellent match with one of the AG
Dra lines (Table~\ref{unid-corr}). Another good wavelength match is
found for the \lam1139.494 transition, however, the predicted
\lam1139/\lam1141 ratio is much lower than the observed ratio. None of
the other transitions can be identified in the AG Dra spectrum.

Applying the same model to \ion{Fe}{iii} yields two lines whose
wavelengths are close to measured lines: \lam1066.195 and
\lam1142.956. The latter is the strongest of the predicted lines in
the \fuse\ waveband and is in excellent wavelength agreement with the
line measured at 1142.429\,\AA. The predicted \lam1070.557 line would
be expected at a wavelength of $\approx$1070.16\,\AA\ which is
coincident with the H$_2$ L3P3 \lam1070.142 interstellar absorption
line and so is not seen. Although the model \lam1066.195 line is
expected to be much weaker than the \lam1070.557 and \lam1142.956
lines, there is a weak line found at a corrected wavelength of
1066.166\,\AA\ which we identify with this transition.

We conclude that two of the unidentified lines can be identified as
\ion{Fe}{ii} \lam1141.172 and \ion{Fe}{iii} \lam1142.429, fluoresced
by \ion{O}{vi} \lam1032. Two more lines can be tentatively identified
as \ion{Fe}{ii} \lam1139.494 and \ion{Fe}{iii} \lam1066.195 based on
wavelength matches, although the theoretically predicted fluxes are
weaker than the observed values. As \ion{Fe}{ii} and \ion{Fe}{iii} are
extremely complex ions, there is the possibility that the atomic
data used in the theoretical model may cause the discrepancy in the
fluxes.

\begin{deluxetable}{llll}
\tablecaption{Predicted fluorescence lines.\label{fluor-flux}}
\tablehead{Pumping line &Pumped transition &Fluoresced transitions
&Relative flux \\
&(\AA) &(\AA) &
           }
\tablewidth{5.4in}
\startdata
\cutinhead{Fluoresced \ion{Fe}{ii} lines}
\ion{O}{vi} \lam1032  &1032.041 &978.697 &6.8 \\
       &&981.468 &5.5 \\
       &&1114.355 &1.8 \\
       &&1126.677 &2.8 \\
       &&1139.494 &1.3 \\
       &&1141.172 &10.0 ($*$)\\
       &&1142.370 &2.8 \\
       &&1776.421 &2.5\\
       &&1776.660 &14.4\\
       &&1881.201 &16.7 \\
       &&1884.116 &11.3 \\
\ion{O}{vi} \lam1037  &1037.808 &1448.393 &1.3 \\
       &&1663.782 &0.7 \\
&\\
\cutinhead{Fluoresced \ion{Fe}{iii} lines}
\ion{O}{vi} \lam1032  &1032.119 & 1066.195 &1.5 \\
       &&1070.557 &9.1\\
       &&1142.956 &10.0 ($*$)\\
       &&2097.696 &30.4 \\
\ion{O}{vi} \lam1037 &1037.459 &1035.769 &1.4 \\
       &&1071.747 &0.6 \\
       &&1142.461 &0.7 \\
       &&2103.809 &1.9 \\
\enddata
\end{deluxetable}

\section{Emission line widths and shifts}\label{line-wid}

\citet{fried83} reported systematic redshifts of resonance lines
relative to intercombination lines from \emph{IUE} spectra of seven
symbiotic stars (including AG Dra), with magnitudes ranging from 10 to
40~\kms. The values given for AG Dra were 12 to
15~\kms. \citet{viotti84} gave a redshift of 9~\kms\ from their
quiescent 
\emph{IUE} spectra, but also noted that the \ion{N}{v}
$\lambda\lambda$1238, 1242 doublet exhibited a larger redshift of
30~\kms\ relative to the intercombination lines.
The first direct measurement of the \ion{O}{vi} lines by
\citet{schmid99} yielded a redshift of 49~\kms\ relative to the
system radial velocity for the $\lambda$1032 line.

The cause of these redshifts has been attributed to the scattering of
line photons in an expanding medium \citep{fried83}, and this also
explains the asymmetries noted in \emph{IUE} spectra of the \ion{N}{v}
lines \citep{miko95} and the \ion{O}{vi} lines found here.

Before discussing the \fuse\ results, the accuracy of the \fuse\
wavelength scale must first be considered. Due to the lack of an on
board calibration lamp, the absolute wavelength calibration of \fuse\ is
good to $\lesssim 70$~\kms\ for observations with the LWRS
aperture due to motions of the target within the aperture, and the
thermally-induced grating rotation\footnote{The FUSE Wavelength
  Calibration White Paper,
  \url{http://fuse.pha.jhu.edu/analysis/calfuse\_wp1.html}.}. The 
relative wavelength calibration within a single spectrum is accurate
to around 5--10~\kms. In order to determine the absolute wavelength
scale of the spectra it is thus necessary to measure features in the
spectra for which the wavelengths are known through other
methods. Typically one uses interstellar absorption lines for which
velocities are known from \emph{IUE} or \emph{HST} spectra. For AG Dra
we use the interstellar velocity measured by \citet{viotti83} of
$-23$~\kms\ from \emph{IUE} spectra using low ionization species. For
each of the \fuse\ spectra 
several interstellar absorption lines are selected and their centroids
are measured through line fitting. If the average velocity of these
lines relative to their rest wavelengths is $v$, then we apply a
correction of $23-v$~\kms\ to that spectrum. The lines selected in
each spectrum were based on the spectra of \citet{barnstedt00}.

The velocities of selected emission lines are presented in
Table~\ref{tbl.shifts}. For the asymmetric \lam1032 line the centroid was
estimated through minimizing the quantity

\begin{equation}
g(\lambda) = |\lambda-\lambda_j|\times  F(\lambda_j)
\end{equation}

\noindent where $F(\lambda_j)$ is the flux in the wavelength bin $\lambda_j$.

\begin{deluxetable}{llll}
\tablecaption{Velocity shifts of \fuse\ lines.\label{tbl.shifts}}
\tablehead{Ion   &Line &Velocity shift &Comment$^a$\\
&(\AA) &(\kms)& }
\tablewidth{5in}
\startdata
\ion{He}{ii} 
      &920.561 &$-155.7$ &L18P2 \lam920.242 (r)\\
      &922.748 &$-133.9$ &\\
      &923.796 &$-153.4$ &\\
      &927.851 &$-145.4$ \\
      &930.342 &$-146.9$ \\
      &937.394 &$-148.7$ &L16P3 \lam936.859 (b)\\
      &942.513 &$-141.5$ &\\
      &949.329 &$-153.5$ &\\
      &958.698 &$-142.9$ &\\
\ion{S}{vi}
      &944.523 &$-134.3$ &\\
\ion{C}{iii}
      &977.020 &$-127.9$ &\ion{O}{i} \lam976.448 (b), \ion{C}{iii}
      \lam977.020 (r)\\ 
\ion{Ne}{vi} 
      &992.731 &$-130.8$ &\\
      &997.169 &$-140.4$ &\\
      &999.291 &$-143.7$ &\\
      &1005.789 &$-138.9$ & L8P2 \lam1005.397\\
\ion{O}{vi}
      &1031.926 &$-109.5$ \\
      &1037.617 &$-102.6$ & \ion{C}{ii} \lam1037.020 (b) \\
\ion{S}{iv}
      &1062.664 &$-127.2$ \\
      &1072.973 &$-124.6$ \\
\ion{Ne}{v}
      &1136.519 &$-136.1$ \\
      &1145.607 &$-137.9$ \\
\enddata
\tablenotetext{a}{Interstellar absorption lines that affect the
  velocity shift determination 
  are identified here. The letter in bracket denotes whether the
  absorption is to the blue (b) or red (r) side of the stellar
  emission line.} 
\end{deluxetable}

The \ion{Ne}{v} and \ion{Ne}{vi} lines are intercombination
transitions, and are found at velocities of $-125$--$-135$~\kms. 
Aside from the
\ion{O}{vi} lines that are affected by P Cygni profiles, the other
resonance lines are from \ion{S}{vi} and \ion{S}{iv}, which are
consistent with the intercombination lines. Thus there is no
indication from the \fuse\ spectra that the resonance lines are
redshifted relative to the intercombination lines. The consistency of
the \ion{S}{vi} velocity shift with the intercombination lines reveals
that this high ionization state is not showing a P Cygni absorption,
restricting the size of the region participating in the wind.

The \ion{He}{ii} Balmer
lines are all significantly blueshifted relative to the other emission
lines, which may indicate that the model assumed in deriving the rest
wavelengths is in error (Appendix~A).

The centroid of the \ion{O}{vi} \lam1032 line is
redshifted relative to the systemic velocity by 39~\kms\ in reasonable
agreement with the value of 49~\kms\ found by \citet{schmid99}.

Accurate measurements of emission line widths require good
signal-to-noise relative to the continuum level, and so we give values
only for the strongest lines in the spectrum in
Table~\ref{tbl.widths}. The \ion{O}{vi} width represents that of the
observed profile. One can estimate the true width of the line by
assuming that it would have a shift comparable to the other emission
lines in the spectrum of $\approx -130$~\kms, leading to a width of
$\approx 105$~\kms. One may speculate that the broader width found for
\ion{Ne}{vi} compared to \ion{Ne}{v} may be due to the \ion{Ne}{vi}
ions participating in the wind seen in the \ion{O}{vi} profile.

\begin{deluxetable}{llllll}
\tablecaption{Widths of selected \fuse\ emission lines.\label{tbl.widths}}
\tablehead{Ion &Wavelength (\AA) &Width (\AA) & Width (\kms)
           }
\tablewidth{3.6in}
\startdata
\ion{He}{ii}  &958  &0.177 &55.4 \\
\ion{Ne}{vi}  &999  &0.257 &77.1 \\
\ion{O}{vi}   &1032 &0.218 &63.3 \\
\ion{Ne}{v}   &1136 &0.192 &50.7 \\
\enddata
\end{deluxetable}

\section{Time variability}\label{sect.time-var}

Including the present observations, the 900--1200\,\AA\ region of the
AG Dra spectrum has now
been observed four times, and the
fluxes of the prominent emission lines are given in
Table~\ref{tbl.phase-flux}. The ORFEUS-I/BEFS fluxes have been derived from
the archived spectra available via the Multimission Archive at STScI
(MAST)\footnote{http://archive.stsci.edu/mast.html}, while the
ORFEUS-II/TUES fluxes are taken from \citet{schmid99}.

Monitoring of AG Dra with \iue\ showed that the emission lines are
modulated by a factor two during the orbit, with the strongest fluxes
at phase 0.5 \citep{griestra99}, when the white dwarf is in front of
the giant. Both ORFEUS measurements of the \ion{O}{vi} and
\ion{He}{ii} lines  lie above the \fuse\ values, even though the 2000
March data were obtained close to phase 0.5. This suggests that the two
ORFEUS measurements were not taken during periods of quiescence, and
this is confirmed by inspection of Fig.~7 of \citet{galis99} where the
U band flux is enhanced above quiescent levels at Julian dates
2\,449\,249 and 2\,450\,409, corresponding to the ORFEUS
observations.

Comparing the line fluxes from the two \fuse\ observations shown in
Table~\ref{tbl.lines} shows that the \ion{O}{vi} lines show the
largest decrease in the 2001~April data-set, being reduced by
25\%. This is consistent with the factor of 2 modulation during the
entire AG Dra orbit found by \citet{griestra99}. The \ion{Ne}{vi}
lines which have a comparable excitation potential to the \ion{O}{vi} lines
show a much smaller decrease in flux of around 10\%. This may indicate
that the changes in the \ion{O}{vi} flux may signal variations of the
optical depth with orbital phase.

There was no variability beyond 2$\sigma$ levels observed during
2~ksec of the \fuse\ observations of 
the \ion{O}{vi} lines, despite there being $\approx 35$~counts s$^{-1}$
in the LiF1A channel.

\begin{deluxetable}{lrrrr}
\tablecaption{Emission line fluxes from ORFEUS, \fuse\label{tbl.phase-flux}}
\tablehead{&\multicolumn{4}{c}{Flux ($\times$ $10^{-14}$~\ecs)}\\
\cline{2-5} 
&ORFEUS-I   &ORFEUS-II  & \multicolumn{2}{c}{\fuse}\\
\cline{4-5}
\hfill Phase: &0.723 &0.833  &0.036  & 0.277\\
Line & }
\tablewidth{4in}
\startdata
\ion{O}{vi} \lam1032  &8610 &6210 &5290 &3930\\
\ion{He}{ii} \lam1085 &536  &248  &240  &229\\
\ion{Ne}{vi} \lam999  &448  &--   &140  &135\\
\ion{Ne}{v}  \lam1136 &34   &--   &27   &18\\
\enddata
\end{deluxetable}

\section{Discussion}\label{sect.discussion}

The previous sections have presented \fuse\ spectra obtained in
2000~March and 2001~April during a quiescent phase of AG Dra's
$\approx$15 year outburst cycle. These represent the first high
resolution, high sensitivity 
spectra obtained in the 905--1187\,\AA\ spectral region for AG
Dra. The high ionization states found in the \fuse\ bandpass reveal
significant new information about the system, unavailable from
earlier, longer wavelength
IUE spectra. In particular, the presence of the 
\ion{Ne}{vii} \lam973 recombination line
implies the existence of \ion{Ne}{viii} in the nebula, the highest
ionization species ever recorded for AG Dra. The \ion{O}{vi} \lam1032
line has a P Cygni profile, demonstrating a wind in the high ionization
part of the nebula during quiescence. Previously P Cygni profiles had
only been seen in IUE spectra obtained during outbursts. The
\ion{Ne}{vi} intercombination lines are sensitive to the radiation
field of the white dwarf and their ratios imply that the \ion{Ne}{vi}
emitting region (and thus also the \ion{O}{vi} emitting region) is at
least 300 white dwarf radii (or 0.14~AU for $R_{\rm wd}\approx
0.10~R_\odot$) from the white dwarf. Evidence for the \ion{O}{vi}
being formed far from the white dwarf is also found from the
\ion{Fe}{ii} and \ion{Fe}{iii} fluorescence lines: the \ion{O}{vi}
resonance lines are capable of fluorescing iron lines in the giant's
atmosphere, whereas \ion{H}{i} Ly\al\ is not, suggesting \ion{O}{vi}
is formed close to the giant's surface, while Ly\al\ is formed much
further out.

The classical view of symbiotic
nebulae has material flowing out from the giant and being ionized in
the vicinity of 
the white dwarf. High ionization species (such as \ion{He}{ii} and
\ion{O}{vi}) are found in the
region around the white dwarf \citep[e.g.][]{fried83}, while low
ionization species are formed at increasingly further distances.
An alternative theory is that the strong nebular lines in high density
systems such as AG Dra are formed in the illuminated part of the
giant's atmosphere, helping to explain the orbital modulation of the
UV line fluxes found from \iue\ \citep[e.g.,][]{altamore81}.
\citet{proga96,proga98} have
created non-LTE photoionization 
models for the ionization of a giant's atmosphere in the case where it
is illuminated by a hot companion star and have demonstrated that the
general features of symbiotic star spectra can be reproduced with high
effective temperatures and luminosities of the companion. The large
emission line fluxes of highly-ionized species can only be reproduced
if the giant has a wind extending to 2--3 giant radii.

\begin{figure}[h]
\epsscale{0.7}
\plotone{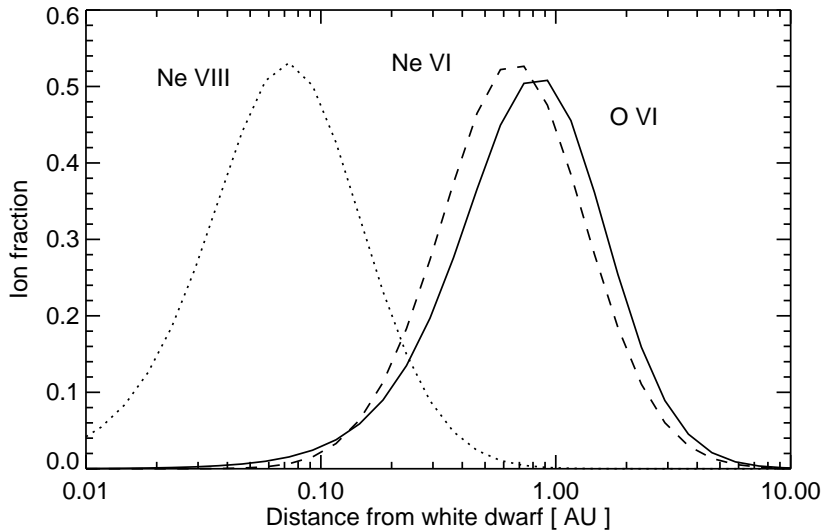}
\caption{Predictions for the location of the \ion{O}{vi},
  \ion{Ne}{vi}, and \ion{Ne}{viii} ions in relation to the white
  dwarf, assuming a simple model for the ionization structure of the
  nebula (App.~\ref{model}), and a white dwarf black body temperature
  of 170,000~K.\label{dist-o6-ne6}}
\end{figure}

\begin{figure}[h]
\epsscale{0.7}
\plotone{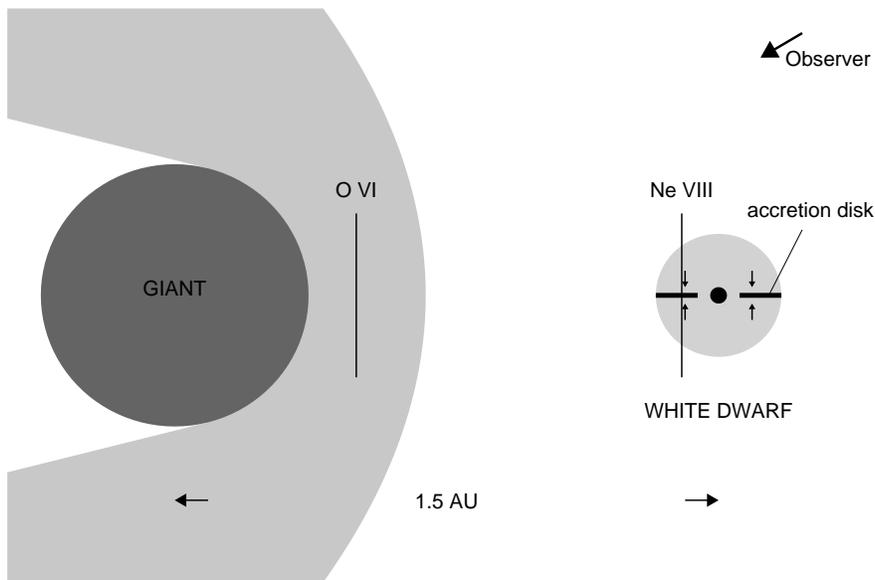}
\caption{A schematic diagram illustrating the locations of the
  \ion{O}{vi} and \ion{Ne}{viii} emitting regions (vertical
  lines) suggested by the present
  work.\label{cartoon}}
\end{figure}

Evidence for the \ion{O}{vi} lines being formed in the giant's
atmosphere comes from the aforementioned \ion{Ne}{vi} result and
Fe fluorescence lines. 
We can also consider
the location of the \ion{Ne}{vi, viii} and \ion{O}{vi} ions in the
system through 
ionization balance arguments. The high ionization potential of these
ions means that they are produced through irradiation from the X-ray
section of the white dwarf's radiation field. Fits to the measured
X-ray spectrum \citep{greiner97} are consistent with a blackbody field
of temperature 170,000~K. By constructing a simple model of the
ionization/recombination processes in the plasma
(Appendix~\ref{model}) we can calculate the fractional
population of \ion{Ne}{vi} and \ion{O}{vi} as a function of distance
from the white dwarf (Fig.~\ref{dist-o6-ne6}). The distance here
corresponds to the inner face of a nebular shell, and the model
assumes that the radiation field is unattenuated between the white
dwarf and inner face. It is clear that such a hot blackbody prevents
\ion{O}{vi} and \ion{Ne}{vi} being formed out to distances of 0.5--1.0
AU. For the model of AG Dra presented in Fig.~9 of \citet{miko95}
this places the \ion{O}{vi} and \ion{Ne}{vi} regions close to the
surface of the giant, and well away from the white
dwarf. \ion{Ne}{viii}, however, must be formed much closer to the
white dwarf. A cartoon illustrating the structure of the AG Dra nebula
is shown in Fig.~\ref{cartoon}. A difficulty for this particular model
lies in the P Cygni profile of the \ion{O}{vi} line, since the
continuum in the far ultraviolet is due to the white dwarf and not the
giant: for white dwarf continuum photons to be absorbed by \ion{O}{vi}
at orbital phase 0.5 (white dwarf in front of the giant), the
\ion{O}{vi} ions are required to lie in front of the white dwarf along
the observer's line of sight. We note however, that the electron
scattered wings around the \ion{O}{vi} lines significantly enhance the
continuum (Fig.~\ref{o6-broad-wings}), and that the P Cygni absorption
extends down to around $3\times 10^{-13}$~\ecs, the level of the white
dwarf continuum, and not to zero flux. We thus suggest that the
\ion{O}{vi} ions are absorbing the electron-scattered \ion{O}{vi}
photons and \emph{not} the stellar continuum. This can be understood if
the electron scattering is occurring in the densest part of the
\ion{O}{vi} region close to the giant, with the absorption of the
scattering wings then occuring in the lower density, extended
wind. The high density of $10^{11}$~cm$^{-3}$ derived from the
analysis of the wings in Sect.~\ref{sect.o6-wings} supports this idea.

A full understanding of the structure of the AG Dra nebula can only
come through detailed modelling of the ionization and wind structure,
constrained by high quality spectral data.

\acknowledgements

We thank S.D.\ Friedman for scheduling the 2000~March
observation. P.R.\ Young thanks W.M.~Moos for the allocation of
observing time for the P248 program.
J.~Aufdenberg and R.L.~Kurucz are thanked for useful discussions.

\appendix

\section{Rest wavelengths of \fuse\ emission lines}

\subsection{Ne\,V}

The two \ion{Ne}{v} lines seen in the \fuse\ spectrum arise from
decays of the 2p $^5$S level to the ground 2s $^3P_{1,2}$
levels. The separation of the two ground levels is strongly
constrained by the measurement of the $^3P_1$ -- $^3P_2$
transition wavelength from SWS 
spectra obtained with ISO by 
\citet{feucht97}. They find a wavelength of
14.3217$\pm$0.0002\,$\mu$m. 

The UV lines have only previously been measured by space-borne, solar UV
spectrometers.
\citet{sandlin} give wavelengths of
1136.51$\pm$0.02\,\AA\ and 1145.61$\pm$0.02\,\AA. \citet{edlen85} made
use of these wavelengths to refine the energy levels of \ion{Ne}{v},
leading to revised wavelengths of 1136.51\,\AA\ and 1145.59\,\AA. More
recently, the SUMER instrument on board SOHO has
measured the lines. \citet{feldman97} give wavelengths of
1136.52\,\AA\ and 1145.62\,\AA\ from off-limb spectra, while
\citet{curdt01} give wavelengths of 
1136.56\,\AA\ and 1145.66\,\AA\ from disk spectra. The off-limb spectra
are hampered by blending with coronal lines, though, while disk
spectra of solar transition region lines typically show redshifts of
5--10~\kms\ \citep[e.g.,][]{peter99}. A detailed reassessment of SUMER
disk spectra (Curdt 2001, 
private communication) yields revised wavelengths of
1136.551$\pm$0.020\,\AA\ and 1145.632$\pm$0.010\,\AA. We correct
these assuming a redshift of 5~\kms\ and use the error estimates from
the UV and IR wavelength measurements to minimize the energy level
separations of the three levels. These yield new wavelengths for the
UV lines of 1136.532\,\AA\ and 1145.615\,\AA\ which are used in the AG
Dra analysis.

\subsection{Ne\,VI}

As with the \ion{Ne}{v} lines, the \ion{Ne}{vi} intercombination lines
have only been measured in solar spectra. The most recent
determination of the lines' wavelengths is that of \citet{dwivedi99}
from off-limb SUMER spectra. However, these wavelengths show a serious
discrepancy with the measurement of the \ion{Ne}{vi} ground transition
wavelength of \citet{feucht97} from the SWS instrument on ISO. In
particular the Dwivedi et al.\ wavelengths for the $^2P_{1/2}$ --
$^4P_{1/2, 3/2}$ transitions imply, when combined with the
\citet{feucht97} wavelength, that the $^2P_{3/2}$ --
$^4P_{1/2, 3/2}$ transitions occur at 1005.777\,\AA\ and
1010.303\,\AA. Dwivedi et al.\ actually measure the latter lines at
wavelengths 1005.696 and 1010.247\,\AA\ -- differences of 0.081\,\AA\
and 0.056\,\AA\ that are much larger than the 0.015\,\AA\ accuracy
quoted by \citet{dwivedi99}.

Analysis of more recent SUMER spectra (W.~Curdt, private
communication, 2001) with updated calibration files has resulted in
revised wavelengths for the \ion{Ne}{vi} lines giving values of
992.731, 997.169, 999.291, 1005.789 and 1010.323\,\AA. These values
have been used in the AG Dra analysis.

\subsection{He\,II}

Each of the \ion{He}{ii} Balmer series lines consists of seven individual
transitions that, in terms of wavelength, can be split into two
groups. E.g., for the n=5 transitions the groups are at average
wavelengths of 1084.912\,\AA\ and 1084.977\,\AA\ (based on data from
the NIST database) -- i.e., a velocity
separation of 18.0~\kms. In order to compare the velocity shifts of
the \ion{He}{ii} lines with those from other ions, it is necessary to
model how the individual line components contribute to the total line
flux.

\citet{clegg99} provides such a model for the \ion{He}{ii} \lam1640 (n=3) and
\lam1215 (n=4) lines, yielding estimates of the line centroids for a range
of temperatures and densities, and for Cases A and B of
\citet{baker}. We extend this model for 5$\le$ n$\le$ 20 with the
method outlined in Sect.~3 of \citet{clegg99}, atomic data from
\citet{storey95}, and $A$-values calculated with the FORTRAN routine
described by \citet{storey91}. Energy level data were obtained from
NIST. For n$\ge$12, energy values were not available for all of the
$^2$P, $^2$D and $^2$S levels. However, the wavelength splittings of
the individual line components are largely determined by the splitting
of the n=2 levels for these high n levels.

To determine the expected wavelengths of the \ion{He}{ii} Balmer
lines for AG Dra, an electron density of $10^{10}$~cm$^{-3}$ and a
temperature of $2\times 10^4$~K were assumed together with Case B. The
emissivities and wavelengths of the seven individual line components
were calculated, and a combined synthetic profile was computed by
assuming lines broadened by 50~\kms. The resulting profile was then
fit with a Gaussian, the centroid of this fit is the wavelength
that is listed in Table~\ref{tbl.lines}.

\section{Revised O\,IV density from IUE data}\label{app.o4}

\begin{figure}[h]
\epsscale{0.6}
\plotone{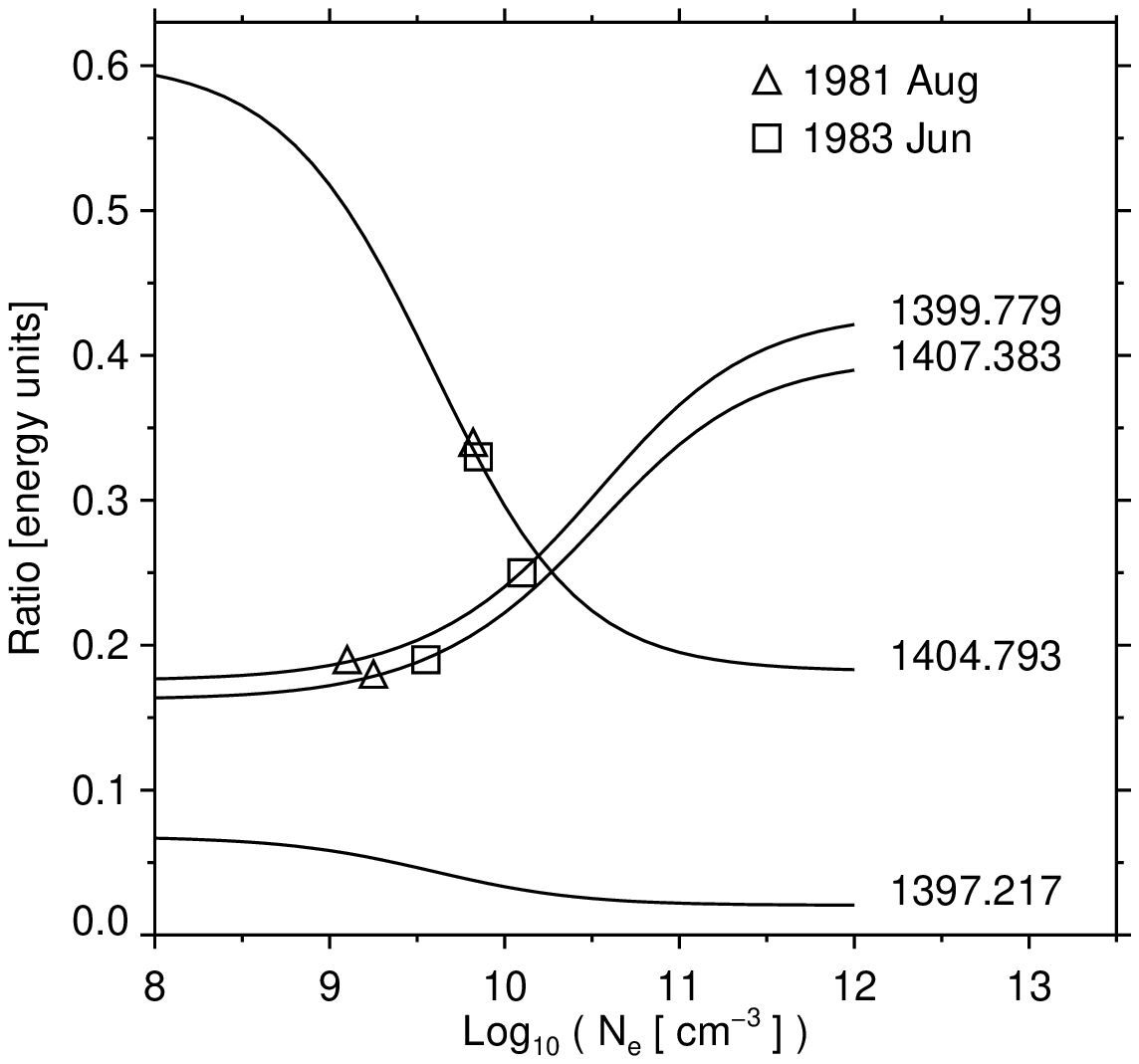}
\caption{Theoretical \ion{O}{iv} line ratios, relative to the
  \lam1401.171 line, from the CHIANTI
database, calculated for an electron temperature of 25\,000~K. The
symbols indicate the ratios from the 1981~August and 1983~June
IUE observations, and the corresponding densities.\label{fig.o4-dens}} 
\end{figure}

\ion{O}{iv} belongs to the same isoelectronic sequence as
\ion{Ne}{vi}, and the intercombination transitions that occur for
\ion{Ne}{vi} in the \fuse\ waveband are found for \ion{O}{iv} between
1397 and 1408\,\AA. These lines form ratios that are sensitive to the
electron density between $10^9$ and $10^{11}$~cm$^{-3}$, and
Fig.~\ref{fig.o4-dens} shows the ratios, relative to the strongest
\lam1401 line, calculated from the CHIANTI
database at an electron temperature of 25\,000~K. Measurements of the
\ion{O}{iv} lines were made by \iue\ and the high resolution mode of
the satellite allowed the individual transitions to be measured. Two
data-sets with good measurements of the lines are from 1981 August 3
and 1983 June 7. The former was obtained several months after the
system went into a major outburst, while the latter was obtained
during a quiescent phase. The line fluxes were very similar in each
case, and the line ratios, relative to \lam1401, are shown in
Table~\ref{tbl.o4-dens} and overplotted on Fig.~\ref{fig.o4-dens}. The
ratios clearly show that the density lies between $10^9$ and
$10^{10}$~cm$^{-3}$. The stronger lines have a higher signal-to-noise
and so we favor densities closer to $10^{10}$~cm$^{-3}$ and this is
the value assumed for the density of the nebula in the present paper.

\begin{deluxetable}{lll}
\tablecaption{IUE \ion{O}{iv} line ratios (energy units)\label{tbl.o4-dens}}
\tablewidth{3.3in}
\tablehead{&SWP14641 &SWP20162 \\
Ratio &1981 August &1983 June}
\startdata
\lam1399/\lam1401   &0.19 &0.25 \\
\lam1404/\lam1401   &0.34 &0.33 \\
\lam1407/\lam1401   &0.18 &0.19 \\
\enddata
\end{deluxetable}

\section{Ionization model}\label{model}

In Sect.~\ref{sect.discussion} results are presented from a model of
the ionization 
balance of oxygen and neon ions in a photoionized plasma. The details
of this model are given here.

The distribution of element ionization states in a plasma photoionized by a
distant radiation source are determined by the distance to the source
and the local density and electron temperature of the plasma. The
number of ions leaving an ionization state $i$ through photoionization
is given by

\begin{equation}
\alpha_i = 4 \pi N_i W(r) \int B(T_*,\lambda) \sigma(\lambda) {\rm d}\lambda
\end{equation}

\noindent where $\sigma$ is the photoionization cross-section,  $B$ is the
specific intensity of the radiation field, and $W(r)$ is the dilution
factor given by

\begin{equation}
W= {1 \over 2} \left[ 1 - \left( 1 - {1 \over r^2} \right)^{1/2} \right]\end{equation}

\noindent and $r=R_*/R$ is the ratio of the radiation source's radius to the
distance of the plasma from the source's center.

The number of ions entering into the state $i$ through recombinations
from the $i+1$ state is

\begin{equation}
N_{\rm e} \beta_i= N_{i+1} N_{\rm e} (\alpha_{\rm rad}(T) + \alpha_{\rm di}(T) )
\end{equation}

\noindent The set of linear equations

\begin{equation}
N_i ( \alpha_i + N_{\rm e} \beta_{i-1}) =
  N_{i-1} \alpha_{i-1} + N_{i+1}N_{\rm e}  \beta_i
\end{equation}

\noindent are solved to yield ion populations as a function of radius

Photoionization cross-sections for both the oxygen and neon ions are
from the analytic fits of \citet{verner95}.
The combined radiative and dielectronic recombination rates for the
oxygen ions are from \citet{nahar99}, 
while for the neon ions the radiative recombination rates are from the
fits provided by D.A.~Verner and available from
http://www.pa.uky.edu/$\sim$verner/fortran.html, while the dielectronic
recombination rates are from \citet{mazz98}.


\begin{thebibliography}{}

\bibitem[Ake et al.(2000)]{ake00}
  Ake, T. B., Dupree, A. K., Young, P. R., et al. 2000,
  \apj, 538, L87

\bibitem[Altamore et al.(1981)]{altamore81}
  Altamore, A., Baratta, G. B., Cassatella, A., \& Friedjung, M. 1981,
  \apj, 245, 630

\bibitem[Aufdenberg(1993)]{aufd93}
  Aufdenberg, J. P. 1993,
  \apj, 87, 337

\bibitem[Baker \& Menzel(1938)]{baker}
  Baker, J. G., \& Menzel, D. H. 1938,
  \apj, 88, 52

\bibitem[Barnstedt et al.(2000)]{barnstedt00}
  Barnstedt, J., Gringel, W., Kappelmann, N., \& Grewing, M. 2000,
  A\&AS, 143, 193

\bibitem[Bernat \& Lambert(1978)]{bernat78}
  Bernat, A. P., \& Lambert, D. L. 1978,
  \pasp, 90, 520

\bibitem[Blair \& Andersson(2001)]{blair01}
  Blair, W. P. \& Andersson, B.-G. 2001,
  The \fuse\ Observer's Guide, ver. 3.0 January 2001. Online.
  Available http://fuse.pha.jhu.edu/support/guide/guide.html

\bibitem[Boyce \& Rieke(1935)]{boyce35}
  Boyce, J. C., \& Rieke, C. A. 1935,
  Phys.\ Rev., 47, 653

\bibitem[Cardelli et al.(1989)]{cardelli}
  Cardelli, J. A., Clayton, G. C., \& Mathis, J. S. 1989,
  \apj, 345, 245

\bibitem[Castor et al.(1970)]{castor70}
  Castor, J. I., Smith, L. F., \& van Blerkom, D. 1970,
  \apj, 159, 1119

\bibitem[Clegg et al.(1999)]{clegg99}
  Clegg, R. E. S., Miller, S., Storey, P. J., \& Kisielius, R. 1999,
  A\&AS, 135, 359

\bibitem[Curdt et al.(2001)]{curdt01}
  Curdt, W., Brekke, P., Feldman, U., Wilhelm, K., Dwivedi, B.N.,
Sch\"uhle, U., \& Lemaire, P. 2001,
  \aap, 375, 591

\bibitem[De Medeiros \& Mayor(1999)]{demed99}
  De Medeiros, J. R., \& Mayor, M. 1999,
  \aaps, 139, 433

\bibitem[Dere et al.(1997)]{dere97}
   Dere, K. P., Landi, E., Mason, H. E., Monsignori-Fossi, B. F., \& 
   Young, P. R. 1997,
   \aaps, 125, 149

\bibitem[Dere et al.(2001)]{dere01}
   Dere, K. P., Landi, E., Young, P. R., \& Del Zanna, G. 2001,
   \apjs, 134, 331

\bibitem[Dwivedi et al.(1999)]{dwivedi99}
  Dwivedi, B. N., Curdt, W., \& Wilhelm, K. 1999,
  \apj, 517, 516

\bibitem[Edl\'en(1983)]{edlen83}
  Edl\'en, B. 1983,
  Phys.\ Scr., 28, 483

\bibitem[Edl\'en(1985)]{edlen85}
  Edl\'en, B. 1985,
  Phys.\ Scr., 31, 345

\bibitem[Ekberg(1993)]{ekberg93}
  Ekberg, J. O. 1993,
  A\&AS, 101, 1

\bibitem[Espey et al.(1996)]{espey96}
  Espey, B., Keenan, F. P., McKenna, F. C., Feibelmann, W. A., \& 
  Aggarwal, K. M. 1996,
  \apj, 465, 965

\bibitem[Fekel et al.(2000)]{fekel00}
  Fekel, F. C., Hinkle, K. H., Joyce, R. R., \& Skrutskie, M. F. 2000,
  \aj, 120, 3255

\bibitem[Feldman et al.(1997)]{feldman97}
  Feldman, U., Behring, W. E., Curdt, W., Sch\"uhle, U., Wilhelm, K.,
  Lemaire, P., \& Moran, T. M. 1997,
  \apjs, 113, 195

\bibitem[Ferland(2000)]{ferland00}
  Ferland, G. J. 2000,
  RMxAC, 9, 153

\bibitem[Feuchtgruber et al.(1997)]{feucht97}
  Feuchtgruber, H., Lutz, D., Beintema, D. A., et al. 1997,
  \apj, 487, 962

\bibitem[Fitzpatrick(1999)]{fitz99}
  Fitzpatrick, E. L. 1999,
  \pasp, 111, 63

\bibitem[Friedjung et al.(1983)]{fried83}
  Friedjung, M., Stencel, R.E., \& Viotti, R. 1983,
  \aap, 126, 407

\bibitem[G\'alis et al.(1999)]{galis99}
  G\'alis, R., Hric, L., Friedjung, M., \& Petr\'\i k, K. 1999,
  \aap, 348, 533

\bibitem[Gonz\'alez-Riestra et al.(1999)]{griestra99}
  Gonz\'alez-Riestra, R., Viotti, R., Iijima, T., \& Greiner, J. 1999,
  \aap, 347, 478

\bibitem[Grevesse \& Sauval(1998)]{grevesse98}
  Grevesse, N., \& Sauval, A. J. 1998,
  Space Science Reviews, 85, 161

\bibitem[Greiner et al.(1997)]{greiner97}
  Greiner, J., Bickert, K., Luthardt, R., Viotti, R., Altamore, A.,
  Gonz\'alez-Riestra, \& Stencel, R. E. 1997,
  \aap, 322, 576

\bibitem[Harper et al.(2001)]{harper01}
  Harper, G. M., Wilkinson, R., Brown, A., Jordan, C., \& Linsky, J. L.
  2001, \apj, 551, 486

\bibitem[Hartman \& Johansson(2000)]{hartman00}
  Hartman, H., \& Johansson, S. 2000,
  \aap, 359, 627

\bibitem[Johansson(1988)]{johan88}
  Johansson, S. 1988,
  \apj, 327, L85

\bibitem[Kafatos et al.(1993)]{kafatos93}
  Kafatos, M., Meier, S. R., \& Martin, I. 1993,
  \apjs, 84, 201

\bibitem[Kaufman \& Martin(1989)]{kauf89}
  Kaufman, V., \& Edl\'en, B. 1989,
  J.\ Opt.\ Soc.\ Am.\ B6, 1769

\bibitem[Kenyon et al.(1993)]{kenyon93}
  Kenyon, S., Miko{\l}ajewska, J., Miko{\l}ajewski, M., Polidan,
  R. S., Slovak, M. H. 1993,
  \aj, 106, 1573

\bibitem[Lutz et al.(1987)]{lutz87}
  Lutz, J. H., Lutz, T. E., Dull, J. D., \& Kolb, D. D. 1987,
  \aj, 94, 463

\bibitem[Mazzotta et al.(1998)]{mazz98}
  Mazzotta, P., Mazzitelli, G., Colafrancesco, S., \& Vittorio,
  N. 1998,
  A\&AS, 133, 403

\bibitem[Meinunger(1979)]{mein79}
  Meinunger, L. 1979,
  IBVS, No.~1611

\bibitem[Mihalas(1970)]{mihalas70}
  Mihalas, D. 1970,
  Stellar Atmospheres, Freeman, San Francisco

\bibitem[Miko{\l}ajewska et al.(1995)]{miko95}
  Miko{\l}ajewska, J., Kenyon, S., Miko{\l}ajewski, M., Garcia, M. R., Polidan,
  R. S. 1995,
  \aj, 109, 1289

\bibitem[Moos et al.(2000)]{moos00}
  Moos, H. W., Cash, W. C., Cowie, L. L., et al. 2000,
  \apj, 538, L1

\bibitem[Morton(2000)]{morton00}
  Morton, D. C. 2000,
  ApJS, in preparation

\bibitem[Nahar(1999)]{nahar99}
  Nahar, S. 1999,
  \apjs, 120, 131

\bibitem[Nussbaumer \& Storey(1984)]{nussb84}
  Nussbaumer, H., \& Storey, P. J. 1984,
  A\&AS, 56, 293

\bibitem[Nussbaumer(1987)]{nussb87}
  Nussbaumer, H. \& Stencel, R. E. 1987,
  Exploring the Universe with the \iue\ Satellite, ed. Y. Kondo, 203
  
\bibitem[Nussbaumer et al.(1995)]{nussb95}
  Nussbaumer, H., Schmutz, W., \& Vogel, M. 1995,
  \aap, 293, L13

\bibitem[Peter \& Judge(1999)]{peter99}
  Peter, H., \& Judge, P. G. 1999,
  \apj, 522, 1148

\bibitem[Proga et al.(1996)]{proga96}
  Proga, D., Kenyon, S. J., Raymond, J. C., \& Miko{\l}ajewska, J. 1996,
  \apj, 471, 930

\bibitem[Proga et al.(1998)]{proga98}
  Proga, D., Kenyon, S. J., \& Raymond, J. C. 1998,
  \apj, 501, 339

\bibitem[Rauch(1997)]{rauch97}
   Rauch, T. 1997,
   \aap, 320, 237

\bibitem[Sandlin et al.(1977)]{sandlin}
  Sandlin, G. D., Brueckner, G. E., \& Tousey, R. 1977,
  \apj, 214, 898

\bibitem[Schmid \& Nussbaumer(1993)]{schmid93}
  Schmid, H. M. \& Nussbaumer, H. 1993,
  \aap, 268, 159

\bibitem[Schmid et al.(1999)]{schmid99}
  Schmid, H. M., Krautter, J., Appenzeller, I., et al. 1999,
  \aap, 348, 950

\bibitem[Smith et al.(1996)]{smith96}
  Smith, V. V., Cunha, K., Jorissen, A., \& Boffin, H. M. J. 1996,
  \aap, 315, 179

\bibitem[Storey \& Hummer(1991)]{storey91}
  Storey, P. J., \& Hummer, D. G. 1991,
  Comp.\ Phys.\ Comm.\ 66, 129

\bibitem[Storey \& Hummer(1995)]{storey95}
  Storey, P. J., \& Hummer, D. G. 1995,
  \mnras, 272, 41

\bibitem[Verner \& Yakovlev(1995)]{verner95}
  Verner, D. A., \& Yakovlev, D. G. 1995,
  \aaps, 109, 125

\bibitem[Viotti et al.(1983)]{viotti83}
  Viotti, R., Ricciardi, O., Ponz, D., et al. 1983,
  \aap, 119, 285

\bibitem[Viotti et al.(1984)]{viotti84}
  Viotti, R., Altamore, A., Baratta, G. B., Cassatella, A., \& Friedjung, M. 1984,
  \apj, 283, 226

\bibitem[Warren et al.(1997)]{warren97}
  Warren, H. P., Mariska, J. T., Wilhelm, K., \& Lemaire, P. 1997,
  \apj, 484, L91 

\bibitem[Young et al.(2003)]{young03}
  Young, P. R., Del Zanna, G., Landi, E., et al. 2003,
  \apjs, 144, 135

\end{thebibliography}
\end{document}